\documentclass[useAMS,usenatbib]{mn2e}

\usepackage[]{natbib}
\usepackage{graphicx}

\title[Transit analysis of CoRoT-targets]{Transit analysis of the CoRoT-5, CoRoT-8, CoRoT-12, CoRoT-18, CoRoT-20, and CoRoT-27 systems with combined ground- and space-based photometry}
\author[St. Raetz et al.]{St. Raetz$^{1,2,3}$\thanks{E-mail:raetz@astro.uni-tuebingen.de}, A. M. Heras$^{3}$, M. Fern\'{a}ndez$^{4}$, V. Casanova$^{4}$, C. Marka$^{5}$ \\
$^{1}$Institute for Astronomy and Astrophysics T\"{u}bingen (IAAT), University of T\"{u}bingen, Sand 1, D-72076 T\"{u}bingen, Germany \\
$^{2}$Freiburg Institute of Advanced Studies (FRIAS), University of Freiburg, Albertstra\ss{}e 19, D-79104 Freiburg, Germany\\
$^{3}$Science Support Office, Directorate of Science, European Space Research and Technology Centre (ESA/ESTEC), Keplerlaan 1,\\ 2201 AZ Noordwijk, The Netherlands\\
$^{4}$Instituto de Astrof\'{\i}sica de Andaluc\'{\i}a, CSIC, Apdo. 3004, 18080 Granada, Spain\\
$^{5}$Instituto Radioastronom\'ia Milim\'{e}trica (IRAM), Avenida Divina Pastora 7, E-18012 Granada, Spain \\
}

\begin{document}

\date{Accepted 2018 November 8. Received: 2018 November 7; in original from 2018 April 6}

\pagerange{\pageref{firstpage}--\pageref{lastpage}} \pubyear{2002}

\maketitle

\label{firstpage}

\begin{abstract}
We have initiated a dedicated project to follow-up with ground-based photometry the transiting planets discovered by \textit{CoRoT} in order to refine the orbital elements, constrain their physical parameters and search for additional bodies in the system. \\ From 2012 September to 2016 December we carried out 16 transit observations of six \textit{CoRoT} planets (CoRoT-5\,b, CoRoT-8\,b, CoRoT-12\,b, CoRoT-18\,b, CoRoT-20\,b, and CoRoT-27\,b) at three observatories located in Germany and Spain. These observations took place between 5 and 9 yr after the planet's discovery, which has allowed us to place stringent constraints on the planetary ephemeris. In five cases we obtained light curves with a deviation of the mid-transit time of up to $\sim$115\,min from the predictions. We refined the ephemeris in all these cases and reduced the uncertainties of the orbital periods by factors between 1.2 and 33. In most cases our determined physical properties for individual systems are in agreement with values reported in previous studies. In one case, CoRoT-27\,b, we could not detect any transit event in the predicted transit window.    

\end{abstract}


\section{Introduction}

\begin{figure}
  \centering
  \includegraphics[width=0.33\textwidth,angle=270]{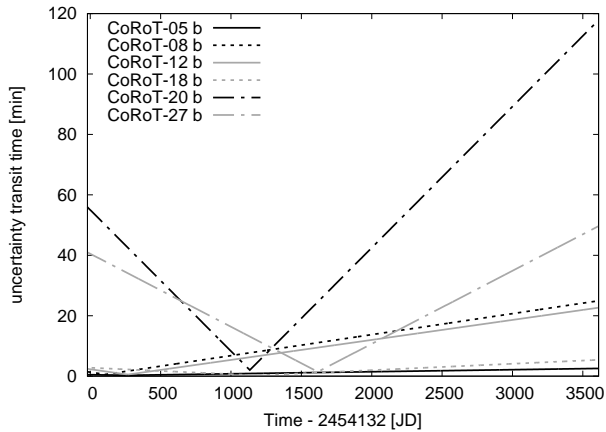}
  \caption{Propagation of the uncertainties on the original published ephemeris for our targets from the start of \textit{CoRoT}'s first science run in 2007 January to our last observation 2016 December. The change points of the lines (close to zero) denotes the transit time at epoch zero (transit discovery).}
  \label{period_err}
\end{figure}

The study of transiting extrasolar planets was revolutionized by the data obtained by space telescopes like \textit{CoRoT} and \textit{Kepler}, as they provide high-precision, high-cadence, continuous light curves (LCs) of a very high number of stars. Thanks to these extraordinary capabilities, the first rocky super-Earths were detected \citep[CoRoT-7b, Kepler-10b,][]{2009A&A...506..303Q,2011ApJ...729...27B}, starting a new era of exoplanet discoveries. \\ \textit{CoRoT} (convection, rotation and planetary transits) was the first space mission dedicated to the detection of transiting planets. The mission was launched in 2006 December and started its first science observation in 2007 January. The spacecraft was equipped with a 27-cm telescope and a 4-CCD wide-field camera. Each pair of CCDs was designed for one of the two main goals of the mission, asteroseismology, or exoplanets. A complete overview on the \textit{CoRoT} mission can be found in `The \textit{CoRoT} Legacy Book: The adventure of the ultra high precision photometry from space' \citep{2016cole.book.....C}. Because of its low-earth orbit, \textit{CoRoT} could point in one direction for not longer than 6 months per year to avoid the Sun entering in its field of view (FoV). The $\sim$6 month observing time in one direction was divided into two separate runs lasting $\sim$30\,d (short run, SR) and $\sim$150\,d (long run, LR). \textit{CoRoT} observed the fields with two cadence modes, a short cadence of 32\,s exposure time and a long cadence with 16 exposures of 32\,s stacked together resulting in 512\,s cadence \citep{2016cole.book...41O}. While most stars were observed in long cadence mode, the 32\,s exposures were only downloaded for selected targets i.e. after the detection of transit like events.\\ In 2009 March, the satellite suffered a loss of communication with one of the data processing units (DPU), which reduced the FoV by 50\%. In 2012 November, the second and last DPU failed resulting in the end of the mission in 2013 June. So far 34 confirmed exoplanets have been published and $\sim$500 candidate exoplanets are awaiting evaluation \citep{2016cole.book.....C}. \\ To truly benefit from \textit{CoRoT}'s planet findings, the planet and orbit parameters need to be accurately determined. Since \textit{CoRoT} could observe transiting planets only for a maximum duration of 150\,d, ground-based follow-up is mandatory to extend the observational baseline. We have therefore initiated a dedicated project to combine the unprecedented precision of \textit{CoRoT} LCs with ground-based follow-up photometry, in order to refine the planets orbital elements, constrain their physical parameters and search for additional bodies in the system. We selected 12 suitable targets that fulfilled the following criteria:
\begin{itemize}
\item The brightness of the host star is V$<$16 mag and the transit depth is at least 8 mmag, to ensure sufficient photometric and timing precision at 1-2 m class ground-based telescopes.
\item The orbit of the known transiting planet is not well constrained through radial velocity (RV) observations or shows non-zero eccentricity (though the circularization time-scale is much shorter than the system age) and/or the data presents deviant RV points, possibly indicating a perturber.
\item Timing errors are critically large, which would impede the planetary transit observations within a few years.
\end{itemize}
A short description and first results of this study were published in \citet[][]{2015EPJWC.10106054R}. Here, we report on our observations of six of these targets, CoRoT-5, CoRoT-8, CoRoT-12, CoRoT-18, CoRoT-20, and CoRoT-27. Table~\ref{Werte_CoRoT} summarizes the literature values of the physical properties of these systems. Fig.~\ref{period_err} gives the propagation of the original published ephemeris uncertainties for these targets to the present. In particular, CoRoT-20 and CoRoT-27 are of special interest as they have the largest uncertainties in our sample. Moreover, both targets are massive hot Jupiters and, hence, very interesting systems to study formation, migration, and evolution of gas giant planets.

\begin{table*}
\centering
\caption{Physical and orbital properties of the observed systems summarized from the literature.}
\label{Werte_CoRoT}
\begin{tabular}{cr@{\,$\pm$\,}lr@{\,$\pm$\,}lr@{\,$\pm$\,}lr@{\,$\pm$\,}l}
\hline \hline
Object & \multicolumn{2}{c}{CoRoT-5} & \multicolumn{2}{c}{CoRoT-8} & \multicolumn{2}{c}{CoRoT-12} & \multicolumn{2}{c}{CoRoT-18}\\ \hline
Epoch zero transit time $T_{0}$  [d] & \multicolumn{2}{c}{2454400.19885 [1]} & \multicolumn{2}{c}{2454239.03311 [2]} & \multicolumn{2}{c}{2454398.62707 [4]} & \multicolumn{2}{c}{2455321.72412 [6]} \\
 & & 0.00020 [1] & & 0.00078 [2] & & 0.00036   [4] & & 0.00018 [6] \\
Orbital period $P$  [d] & 4.0378962 & 0.0000019 [2] & 6.212381 & 0.000057 [2] & 2.828042 & 0.000013  [4] & 1.9000693 & 0.0000028 [6] \\
Semimajor axis $a$ [au] & 0.05004 & 0.001265 [2] & 0.063 & 0.001 [3] & 0.04016 &  $^{0.00093}_{0.00092}$  [4] & 0.02860 & 0.00065 [7] \\
Inclination $i$  [$^{\circ}$] & 86.24 & 0.53 [2] & 88.4 & 0.1 [3] & 85.79 & 0.43  [2] & 86.5 & $^{1.4}_{0.9}$  [6] \\
Eccentricity $e$ & 0.09 & $^{0.09}_{0.04}$ [1]  & \multicolumn{2}{c}{0* [3]} & 0.070 & $^{0.063}_{0.042}$  [4] & 0.10 & 0.04 [8] \\
Mass star $M_{\mathrm{A}}$  [M$_{\odot}$] & 1.00 & 0.02 [1] & 0.88 & 0.04 [3] & 1.078 &  $^{0.077}_{0.072}$   [4] & 0.861 & 0.059 [7] \\
Radius star $R_{\mathrm{A}}$  [R$_{\odot}$] & 1.186 & 0.040 [1] & 0.77 & 0.02 [3] & 1.046 & 0.042  [2] & 0.924 & 0.057 [7] \\
Effective temperature $T_{\mathrm{eff}}$  [K] & 6100 & 65 [1] & 5080 & 80 [3] & 5675 & 80  [4] & 5440 & 100 [6] \\
Surface gravity star log$\,g_{\mathrm{A}}$ & 4.19 & 0.03 [1] & 4.58 & 0.08 [3] & 4.375 & $^{0.065}_{0.062}$  [4] & 4.442 & 0.043 [7] \\
Metallicity $\left[ \frac{Fe}{H}\right] $ & -0.25 &  0.06 [1] & 0.3 &  0.1 [3] & 0.16 &  0.10  [4] & -0.1 & 0.1 [6] \\
Mass planet $M_{\mathrm{b}}$  [$M_{\mathrm{Jup}}$] & 0.467 & $^{0.047}_{0.024}$ [1] & 0.22 & 0.03 [3] & 0.917 & $^{0.070}_{0.065}$  [4] & 3.27 & 0.17 [7] \\
Radius planet $R_{\mathrm{b}}$  [$R_{\mathrm{Jup}}$] & 1.388 & $^{0.046}_{0.047}$ [1] & 0.57 & 0.02 [3] & 1.350 & 0.074  [2] & 1.251 & 0.083 [7] \\
Distance [pc] & \multicolumn{2}{c}{}  & 380 & 30 [3] & 1150 & 85 [4] & 870 & 90 [6] \\
Age [Gyr] & \multicolumn{2}{c}{$\sim$\,5.5\,-\,8.3 [1]} & \multicolumn{2}{c}{$\leq$\,3 [3]} & 6.3 & 3.1  [4] & 0.1 & $^{0.8}_{0.04}$ [6] \\
Spectral type &  \multicolumn{2}{c}{F9V [1]} &  \multicolumn{2}{c}{K1V [3]} &  \multicolumn{2}{c}{G2V [5]} &  \multicolumn{2}{c}{G9V [6]} \\
RA & \multicolumn{2}{c}{06h45m07s [1]} & \multicolumn{2}{c}{19h26m21s  [3]} & \multicolumn{2}{c}{06h43m04s [4]}  & \multicolumn{2}{c}{06h32m41.36s [6]} \\
Dec & \multicolumn{2}{c}{00$^{\circ}$48'55'' [1]} & \multicolumn{2}{c}{01$^{\circ}$25'36'' [3]} & \multicolumn{2}{c}{−01$^{\circ}$17'47'' [4]}  & \multicolumn{2}{c}{-00$^{\circ}$01'53.71'' [6]}\\
$V$ [mag] &  \multicolumn{2}{c}{14.0 [1]}  &  \multicolumn{2}{c}{14.8 [3]} & 15.515 & 0.052 [4] &  15.00 & 0.10 [6]\\
\hline \hline
Object & \multicolumn{2}{c}{CoRoT-20} & \multicolumn{2}{c}{CoRoT-27} & \multicolumn{2}{c}{} & \multicolumn{2}{c}{}\\ \hline
Epoch zero transit time $T_{0}$  [d] &  \multicolumn{2}{c}{2455266.0001 [9]} & \multicolumn{2}{c}{2455748.684 [10]} & \multicolumn{2}{c}{} & \multicolumn{2}{c}{}\\
 & & 0.0014 [9] & & 0.001 [10] & \multicolumn{2}{c}{} & \multicolumn{2}{c}{}\\
Orbital period $P$  [d] & 9.24285 & 0.00030 [9] & 3.57532 &  0.00006 [10] & \multicolumn{2}{c}{} & \multicolumn{2}{c}{}\\
Semimajor axis $a$ [au] & 0.0902 & 0.0021 [9] & 0.0476 & 0.0066 [10] & \multicolumn{2}{c}{} & \multicolumn{2}{c}{}\\
Inclination $i$  [$^{\circ}$] & 88.21 & 0.53  [9]& 86.7 & $^{1.2}_{0.8}$ [10] & \multicolumn{2}{c}{} & \multicolumn{2}{c}{}\\
Eccentricity $e$ & 0.562 & 0.013 [9]& \multicolumn{2}{c}{$<$0.065 [10]} & \multicolumn{2}{c}{} & \multicolumn{2}{c}{}\\
Mass star $M_{\mathrm{A}}$  [M$_{\odot}$] & 1.11 & 0.01 [7] & 1.05 &  0.11 [10] & \multicolumn{2}{c}{} & \multicolumn{2}{c}{}\\
Radius star $R_{\mathrm{A}}$  [R$_{\odot}$] & 1.02 & 0.05 [9] & 1.08 & $^{0.18}_{0.06}$ [10] & \multicolumn{2}{c}{} & \multicolumn{2}{c}{}\\
Effective temperature $T_{\mathrm{eff}}$  [K] & 5880 & 90 [9] & 5900 & 120 [10] & \multicolumn{2}{c}{} & \multicolumn{2}{c}{}\\
Surface gravity star log$\,g_{\mathrm{A}}$ & 4.20 & 0.15 [9] & 4.4 & 0.1 [10] & \multicolumn{2}{c}{} & \multicolumn{2}{c}{}\\
Metallicity $\left[ \frac{Fe}{H}\right] $ & 0.14 & 0.12 [9] & −0.1 & 0.1 [10] & \multicolumn{2}{c}{} & \multicolumn{2}{c}{}\\
Mass planet $M_{\mathrm{b}}$  [$M_{\mathrm{Jup}}$] & 4.24 & 0.23 [9] & 10.39 & 0.55 [10] & \multicolumn{2}{c}{} & \multicolumn{2}{c}{}\\
Radius planet $R_{\mathrm{b}}$  [$R_{\mathrm{Jup}}$] & 0.84 & 0.04 [9] & 1.007 & 0.044 [10] & \multicolumn{2}{c}{} & \multicolumn{2}{c}{}\\
Distance [pc] & 1230 & 120 [9]& \multicolumn{2}{c}{} & \multicolumn{2}{c}{} & \multicolumn{2}{c}{}\\
Age [Gyr] & 0.1 & $^{0.8}_{0.04}$ [9] & 4.21 & 2.72 [10] & \multicolumn{2}{c}{} & \multicolumn{2}{c}{}\\
Spectral type & \multicolumn{2}{c}{G2V [9]}& \multicolumn{2}{c}{G2V [10]} & \multicolumn{2}{c}{} & \multicolumn{2}{c}{}\\
RA & \multicolumn{2}{c}{06h30m53s [6]} & \multicolumn{2}{c}{18h33m59s [10]} & \multicolumn{2}{c}{} & \multicolumn{2}{c}{}\\
Dec & \multicolumn{2}{c}{00$^{\circ}$13'37'' [9]} & \multicolumn{2}{c}{05$^{\circ}$32'18.503'' [10]} & \multicolumn{2}{c}{} & \multicolumn{2}{c}{}\\
$V$ [mag] & \multicolumn{2}{c}{14.66 [9]}&  \multicolumn{2}{c}{15.540 [10]} & \multicolumn{2}{c}{} & \multicolumn{2}{c}{}\\
\hline \hline
\end{tabular}
\\ References: [1] \citet{2009A&A...506..281R}, [2] \citet{2011MNRAS.417.2166S}, [3] \citet{2010A&A...520A..66B}, [4] \citet{2010A&A...520A..97G}, [5] \citet{2011A&A...529A.136E}, [6] \citet{2011A&A...533A.130H}, [7] \citet{2012MNRAS.426.1291S}, [8] \citet{2013A&A...550A..67P}, [9] \citet{2012A&A...538A.145D}, [10] \citet{2014A&A...562A.140P}\\
$^{\ast}$Fixed in radial velocity analysis.
\end{table*}

\section{Observation, data reduction, and photometry}
\label{photometry}

\begin{table*}
\caption{Observatories and instruments used to observe transits of the \textit{CoRoT} targets.}
\label{CCD_Kameras}
\begin{tabular}{ccccccccc}
\hline \hline
Observatory & Long. (E) & Lat. (N) & Altitude & Mirror $\diameter$ & Camera & \# Pixel & Pixel scale & FoV\\ 
& [$^{\circ}$] & [$^{\circ}$] & [m] & [m] & & &  [$''/pix$] & [$'$]\\ \hline
Jena/Germany  & 11.5 & 50.9 & 370 & 0.90$^{a}$ & E2V CCD42-10  & 2048\,x\,2048 & 1.55 & 52.8\,x\,52.8 \\
  &  &  &  &  & (STK)$^{b}$  &  &  &  \\
Sierra Nevada/Spain & 356.6 & 30.1 & 2896 & 1.50 & VersArray:2048B & 2048\,x\,2048 & 0.23 & 7.8\,x\,7.8  \\
Teide/Tenerife & 343.5 & 28.3 & 2390 & 1.00 & Roper Spec Camera & 2048\,x\,2048 & 0.40 & 13.8\,x\,13.8  \\
\hline \hline
\end{tabular}
\\
$^{a}$0.60\,m in Schmidt mode, $^{b}$ \citet{2010AN....331..449M}
\end{table*}

We started our follow-up campaign in 2013 October after first test observations that were carried out in 2012 September. In total, we collected 16 high-precision LCs of the six selected targets, CoRoT-5, CoRoT-8, CoRoT-12, CoRoT-18, CoRoT-20, and CoRoT-27, from 2012 September to 2016 December. Our ground-based observations were performed with three 1-m class telescopes located in Germany and Spain. Summaries of the participating observatories and observations are given in Table~\ref{CCD_Kameras} and Table~\ref{Beobachtungslog}, respectively. We have also re-analysed the \textit{CoRoT} observations for these targets. The details are given in the following sections.  

\subsection{\textit{CoRoT} observations}

We downloaded the fully reduced LCs (N2 -- the primary scientific, Version 2.1 or 2.2)  produced by the \textit{CoRoT} pipeline \citep{2009A&A...506..411A} from the CoRoT archive mirror at the `NASA Exoplanet Archive' \citep[][http://exoplanetarchive.ipac.caltech.edu/]{2013PASP..125..989A}. In all cases we used the white-light LCs. For most of the targets, the LCs consist of the long cadence data at the beginning of the observations as well as short cadence data for the rest. \\ In preparation for the LC analysis we applied several steps to clean the LCs. First, we removed all flagged measurements \citep[flagged e.g. because of energetic particles, South Atlantic Anomaly crossings, Earth eclipses;][]{2016cole.book...61C}. Then we extracted the transits from the LC by using all data points $\pm$0.2\,d around the expected transit time calculated with the published ephemeris. In the same step, we corrected the time stamp, which is given in heliocentric julian date at the end of the measurements in the original LCs, to the middle of the exposure. In step three, we normalized the LCs. After the division by the average out-of-transit flux, additional light (`third' light) $L_{\mathrm{3}}$ induced by contaminants in the aperture around the target star was subtracted from the normalized flux before re-normalizing. In a last step we cleaned the LCs from outliers. By using a moving average of the time-series we created a smoothed LC. Finally, we removed all data points that deviated more than 3$\sigma$ from this smoothed LC.

\subsection{Ground-based observations}

One observation in 2012 was carried out using the `Schmidt Teleskop Kamera' \citep[STK,][]{2010AN....331..449M} mounted at the 90\,cm Schmidt telescope (60\,cm in Schmidt mode) at the University Observatory Jena. With 2048\,$\times$\,2048 pixels and a pixel scale of 1.55\,arcsec/pixel, we could observe a large FoV of 53\,x\,53\,arcmin. \\ Most of the LCs (12 out of 16) were collected with the 1.5-m reflector at the Observatorio de Sierra Nevada (OSN), which is operated by the Instituto de Astrof\'{i}sica de Andaluc\'{i}a, CSIC, Spain. Using a VersArray:2048B CCD camera  (2048\,$\times$\,2048 pixels, pixel scale 0.23\,arcsec/pixel) we covered a FoV of $7.85\,\times\,7.85$\,arcmin. \\ From 2015 November to 2016 June we obtained three additional LCs at ESA's Optical Ground Station (OGS), a 1-m telescope located at the Observatorio del Teide on Tenerife. The mounted spectrograph \citep{2014DPS....4621405S} was used in imaging mode for the observations. The Roper Spec Camera provides 2048\,$\times$\,2048 pixels with a pixel scale of 0.403\,arsec/pixel. The initial FoV of $13.8\,\times\,13.8$\,arcmin was windowed to shorten read-out time.\\ Since the \textit{CoRoT} targets are relatively faint ($V\sim14-15.5\,$mag, see Table~\ref{Werte_CoRoT}) all observations were carried out either in $R$-band or without any filter, with exposure times between 90 and 180\,s. \\ Data reduction and photometry were performed following the procedures described by us in e.g. \citet{2014MNRAS.444.1351R,2016MNRAS.460.2834R}. In short, we subtracted a bias (as overscan for the data of the STK) and a dark frame (only for STK) and divided by a sky flat field using the \begin{scriptsize}IRAF\end{scriptsize}\footnote{\begin{scriptsize}IRAF\end{scriptsize} is distributed by the National Optical Astronomy Observatories, which are operated by the Association of Universities for Research in Astronomy, Inc., under cooperative agreement with the National Science Foundation.} routines \textit{zerocombine}, \textit{darkcombine}, \textit{flatcombine}, and \textit{ccdproc}. For the aperture photometry with 10 different aperture radii we used a script based on the standard \begin{scriptsize}IRAF\end{scriptsize} routine \textit{phot}. Finally, we derived differential magnitudes using an optimized artificial comparison star \citep{2005AN....326..134B}. We chose the aperture radius that produced the lowest LC scatter (lowest standard deviation) for a sample of constant stars. \\ As preparation for the LC analysis we applied part of the LC treatment as explained for the \textit{CoRoT} LCs. In particular, steps three and four were performed to transform the differential magnitudes into fluxes, normalize the LCs, and remove outliers.

\begin{table}
\caption{Summary of our observations at the University Observatory Jena with the STK, the Observatorio de Sierra Nevada (OSN), and ESA's Optical Ground Station (OGS) in the period from 2012 September to 2016 December. $N_{\mathrm{exp}}$: number of exposures, $T_{\mathrm{exp}}$: exposure times}
\label{Beobachtungslog}
\begin{tabular}{lcccc}
\hline \hline
Date & Telescope & Filter & $N_{\mathrm{exp}}$ & $T_{\mathrm{exp}}$  \\
& & &  & [s]  \\\hline
\multicolumn{5}{c}{CoRoT-5}\\\hline
2014 Jan 07 & OSN & $R$ & 132 & 120 \\
2015 Oct 27 & OSN & $R$ & 115 & 150 \\
2016 Dec 20 & OSN & $R$ & 123 & 120 \\\hline
\multicolumn{5}{c}{CoRoT-8}\\\hline
2012 Sep 06 & STK & $R$ & 96 & 120 \\
2016 Jun 16 & OGS & white light & 111 & 180,120 \\\hline
\multicolumn{5}{c}{CoRoT-12}\\\hline
2014 Dec 22 & OSN & $R$ & 116 & 120,150 \\
2015 Nov 15 & OGS & white light & 114 & 120 \\
2016 Feb 25 & OSN & $R$ & 83 & 180 \\\hline
\multicolumn{5}{c}{CoRoT-18}\\\hline
2014 Jan 20 & OSN & $R$ & 101 & 120, 150 \\
2014 Oct 28 & OSN & $R$ &  98 & 150, 160 \\
2014 Nov 16 & OSN & $R$ & 123 & 140 \\
2016 Jan 31 & OSN & $R$ & 112 & 120, 130, 150 \\\hline
\multicolumn{5}{c}{CoRoT-20}\\\hline
2015 Jan 08 & OSN & $R$ & 103 & 150,140 \\
2015 Nov 18 & OGS & white light & 65 & 90 \\\hline
\multicolumn{5}{c}{CoRoT-27}\\\hline
2016 Jun 03 & OSN & $R$ & 126 & 180 \\
2016 Jun 28 & OSN & $R$ & 100 & 180 \\
\hline \hline
\end{tabular}

\end{table}

\section{Light curve analysis}
\label{Transit_Fitting}

The LC analysis was performed by fitting the transit model of \citet{2002ApJ...580L.171M} to the LCs using the Transit Analysis Package\footnote{http://ifa.hawaii.edu/users/zgazak/IfA/\begin{scriptsize}TAP\end{scriptsize}.html} \citep[\begin{scriptsize}TAP\end{scriptsize} v2.1,][]{2012AdAst2012E..30G}. \begin{scriptsize}TAP\end{scriptsize} fits the LCs using \begin{scriptsize}EXOFAST\end{scriptsize} \citep{2013PASP..125...83E} and estimates parameter uncertainties with the wavelet-based technique of \citet{2009ApJ...704...51C}.\\ All \textit{CoRoT} and ground-based LCs of a given target were simultaneously fitted using 10 Markov Chain Monte Carlo (MCMC) chains with $10^{5}$ steps each. The wavelength independent parameters (orbital inclination $i$ and the semimajor-axis scaled by stellar radius $\frac{a}{R_{\mathrm{A}}}$) and the wavelength dependent parameters (planetary to stellar radii ratio $\frac{R_{\mathrm{b}}}{R_{\mathrm{A}}}$ and the limb darkening coefficients) were connected for all LCs and for LCs in the same filter, respectively. The signal-to-noise ratio of the individual LCs was not sufficient (between $\sim$2 and 16) to derive the limb darkening coefficients from the LC analysis \citep{2013A&A...549A...9C}. When the coefficients were allowed to vary in a wide range, the fitting procedure sometimes gave unphysical results. However, to not underestimate the parameter uncertainties by using limb darkening coefficients that were held fixed \citep[see e.g.][]{2013AJ....146..147M}, they were allowed to vary $\pm0.1$ around the theoretical values for the quadratic limb darkening law (used by \begin{scriptsize}TAP\end{scriptsize}). The limb darkening coefficients were inferred from the tables by \citet{2000A&A...363.1081C} and \citet{2010A&A...510A..21S} for the ground-based and the \textit{CoRoT} observations, respectively. Photometric trends in the LCs were fitted simultaneously with the transit. The system parameters resulting from the LC modelling are given in Table~\ref{tbl:TAPmcmc1}.

\begin{table*}
\centering
\caption{System parameters resulting from the simultaneous wavelet-based red noise MCMC analysis of all \textit{CoRoT} and ground-based LCs.}
\label{tbl:TAPmcmc1}
\renewcommand{\arraystretch}{1.1}
\begin{tabular}{lccccc}
\hline \hline
Object & CoRoT-5 & CoRoT-8 & CoRoT-12 & CoRoT-18 & CoRoT-20\\ \hline
    Inclination [$^{\circ}$] & 85.68 $^{+0.18}_{-0.17}$ & 86.88 $^{+0.41}_{-0.34}$ & 85.71 $^{+0.39}_{-0.36}$ & 89.9 $^{+1.6}_{-1.6}$ & 85.9 $^{+2.5}_{-2.2}$ \\
           $a$/$R_{\mathrm{A}}$ & 9.54 $^{+0.20}_{-0.19}$ & 13.7 $^{+1.0}_{-0.8}$ & 8.02 $^{+0.26}_{-0.24}$ & 7.013 $^{+0.078}_{-0.160}$ & 16.5 $^{+2.0}_{-2.7}$ \\
          $R_{\mathrm{b}}$/$R_{\mathrm{A}}$ (\textit{CoRoT} white light) & 0.1155 $^{+0.00083}_{-0.00084}$ & 0.0849 $^{+0.0020}_{-0.0022}$ & 0.1314 $^{+0.0015}_{-0.0015}$ & 0.1331 $^{+0.0014}_{-0.0013}$ & 0.0884 $^{+0.0045}_{-0.0035}$ \\
          $R_{\mathrm{b}}$/$R_{\mathrm{A}}$ (\textit{R}-band) & 0.1123 $^{+0.0022}_{-0.0022}$ & 0.081 $^{+0.011}_{-0.008}$ & 0.1297 $^{+0.0032}_{-0.0033}$ & 0.1410 $^{+0.0020}_{-0.0019}$ & 0.0885 $^{+0.0066}_{-0.0065}$ \\
          $R_{\mathrm{b}}$/$R_{\mathrm{A}}$ (white light) &  & 0.0757 $^{+0.0072}_{-0.0040}$ & 0.1437 $^{+0.0039}_{-0.0042}$ & &\\          
      Linear LD* (\textit{CoRoT} white light) & 0.360 $^{+0.017}_{-0.017}$ & 0.579 $^{+0.020}_{-0.021}$  & 0.472 $^{+0.015}_{-0.014}$ & 0.492 $^{+0.025}_{-0.025}$ & 0.420 $^{+0.050}_{-0.049}$ \\
        Quad LD* (\textit{CoRoT} white light) & 0.271 $^{+0.018}_{-0.018}$ & 0.129 $^{+0.021}_{-0.021}$  & 0.211 $^{+0.015}_{-0.014}$ & 0.199 $^{+0.026}_{-0.026}$ & 0.239 $^{+0.050}_{-0.050}$ \\      
      Linear LD* (\textit{R}-band) & 0.294 $^{+0.052}_{-0.052}$ & 0.502 $^{+0.095}_{-0.099}$ & 0.372 $^{+0.062}_{-0.061}$ & 0.384 $^{+0.041}_{-0.041}$ & 0.298 $^{+0.093}_{-0.094}$ \\
        Quad LD* (\textit{R}-band) & 0.398 $^{+0.053}_{-0.053}$ & 0.227 $^{+0.097}_{-0.098}$ & 0.322 $^{+0.065}_{-0.066}$ & 0.292 $^{+0.047}_{-0.048}$ & 0.335 $^{+0.096}_{-0.098}$ \\
      Linear LD* (white light) &   & 0.36  $^{+0.10}_{-0.10}$   & 0.321 $^{+0.084}_{-0.085}$  & & \\
        Quad LD* (white light) &   & 0.27  $^{+0.10}_{-0.10}$   & 0.297 $^{+0.091}_{-0.090}$ & & \\   
 \hline\hline
\end{tabular}
\\$^{\ast}$ were allowed to vary $\pm0.1$ around the theoretical values (see text)
\end{table*}

\section{Physical properties}

From the system parameters we obtained from the LC modelling, we calculated the physical properties for each of the observed systems. As explained by us in detail in e.g. \citet[][]{2015MNRAS.451.4139R}, we followed the procedures of \citet{2009MNRAS.394..272S}. In a first step, we determined the stellar parameters mass $M_{\mathrm{A}}$, luminosity $L_{\mathrm{A}}$, and age by employing PARSEC isochrones \citep[version 1.2S,][]{2012MNRAS.427..127B}. For transiting planetary systems, a modified version of the Hertzsprung--Russel diagram (HRD) can be drawn by using the mean stellar density $\rho_{\mathrm{A}}$, which can accurately be determined from the LC modelling as shown by \citet{2010arXiv1001.2010W}. The improved orbital period $P$ necessary to calculate $\rho_{\mathrm{A}}$ was derived from all available transit times (see the next section). In addition, we deduced the stellar radius $R_{\mathrm{A}}$ and the surface gravity $g_{\mathrm{A}}$ from the fitted parameters $a$/$R_{\mathrm{A}}$ and $R_{\mathrm{b}}$/$R_{\mathrm{A}}$ and the simplified formulas given in \citet{2009MNRAS.394..272S}, respectively. The planetary parameters $R_{\mathrm{b}}$ and $g_{\mathrm{b}}$ were derived along with the stellar ones. In the next step, we re-determined the planetary mass $M_{\mathrm{b}}$ and computed the planetary density $\rho_{\mathrm{b}}$. We then calculated the planet's equilibrium temperature, $T_{\mathrm{eq}}$, assuming the effective temperature of the host star from the literature and the Safronov number $\Theta$ \citep[a measure of the efficiency with which a planet gravitationally scatters other bodies,][]{1972epcf.book.....S}. Finally we calculated the geometrical parameters, semimajor axis $a$ using Kepler's third law, and the impact parameter $b$.

\section{Transit Timing}

The mid-transit times of each transit were obtained by the simultaneous modelling with \begin{scriptsize}TAP\end{scriptsize} where $T_{\mathrm{c}}$ was always a free parameter. The transit times, which are given in heliocentric julian date and julian date for the \textit{CoRoT} and the ground-based observations, respectively, were converted into the barycentric Julian Date based on the barycentric dynamic time (BJD$_{\mathrm{TDB}}$) using the online converter\footnote{http://astroutils.astronomy.ohio-state.edu/time/utc2bjd.html} by \citet{2010PASP..122..935E}. Our observations were carried out 5-9\,yr, 7\,yr on average, after the \textit{CoRoT} planet discovery. We used the ephemeris that are available in the literature to compute `observed minus calculated' (O--C) residuals for all transit times. As expected from the uncertainties of the published ephemeris, in most cases the observed transit times deviate significantly from the predicted ones. Hence, a re-calculation of the transit ephemeris was necessary for all observed targets. For an exact determination of the ephemeris we plotted the mid-transit times over the epoch and performed a weighted linear fit. \\ Finally, we computed the generalized Lomb--Scargle periodogram \citep[\begin{scriptsize}GLS\end{scriptsize};][]{2009A&A...496..577Z} to search for a periodicity in the transit timing residuals.

\section{CoRoT-5}

\begin{figure*}
 \includegraphics[width=0.235\textwidth, angle=270]{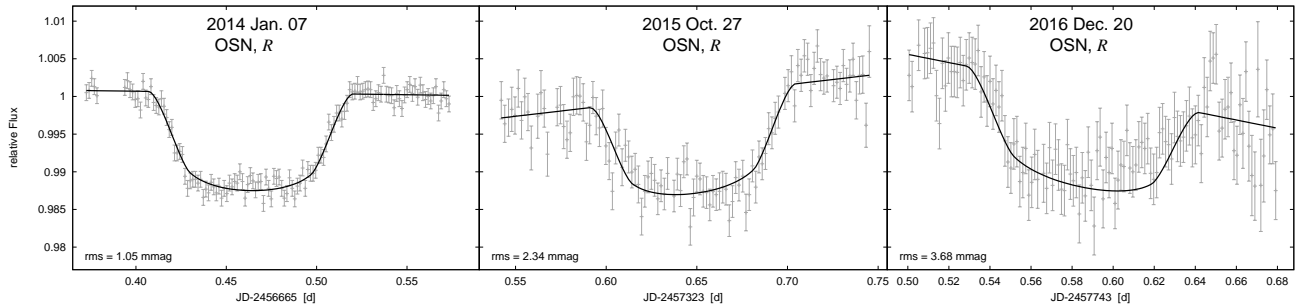}
  \caption{Light curves of CoRoT-5\,b with best-fitting model resulting from the simultaneous fit of all \textit{CoRoT} and ground-based LCs. The dates of observation, observatory, filter, and the $rms$ of the fit are indicated in each individual panel.}
\label{LC_CoRoT5}
\end{figure*}

\begin{figure}
\begin{minipage}[]{0.45\textwidth}
   \includegraphics[width=0.65\textwidth,angle=270]{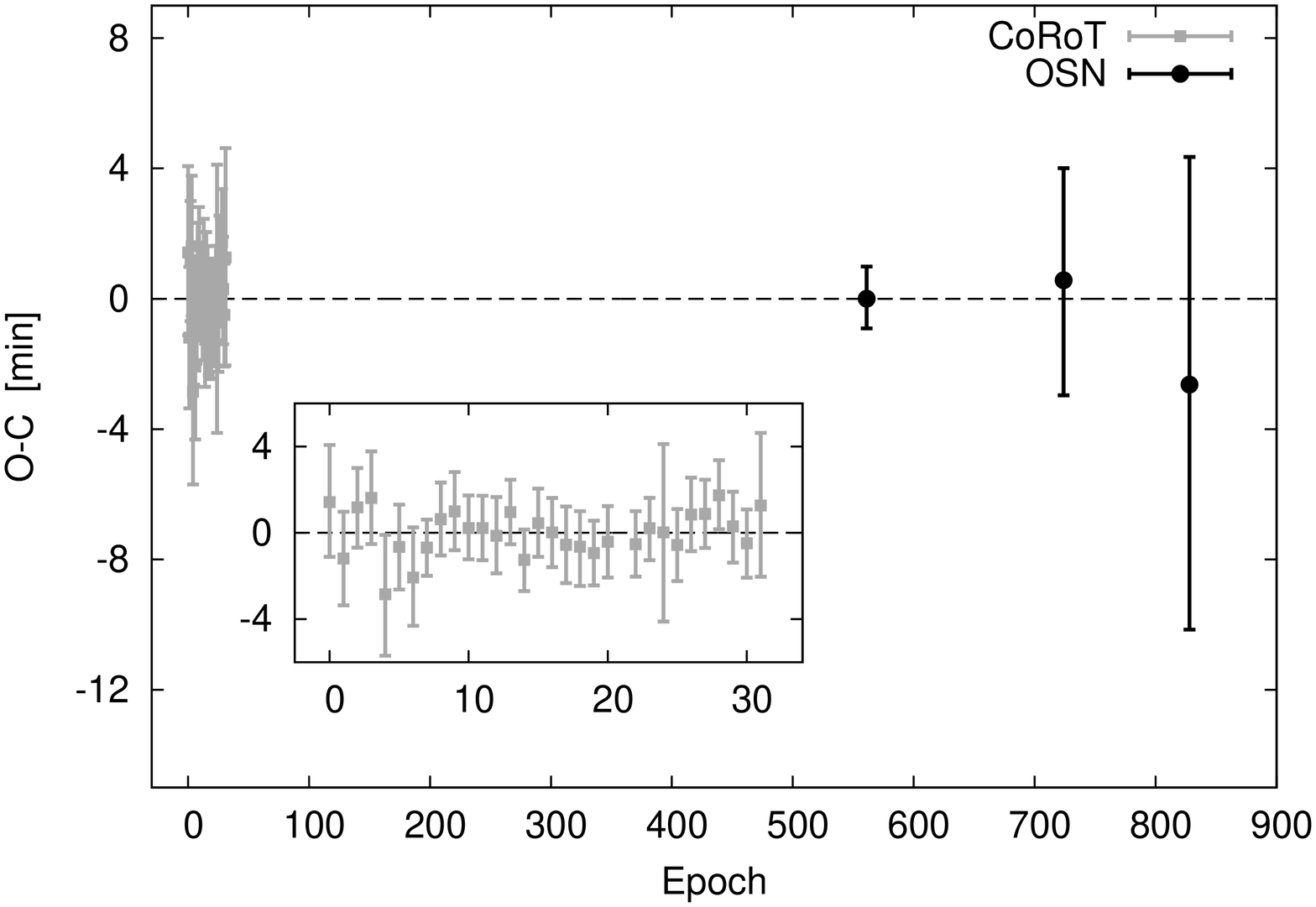}
  \caption{The O--C diagram of CoRoT-5\,b. The grey squares and the black circles denote the \textit{CoRoT} and the OSN transits, respectively. The dashed line represents the updated ephemeris given in equation \ref{Elemente_CoRoT5}. }
  \label{O_C_CoRoT5}
\end{minipage}
\begin{minipage}[]{0.45\textwidth}
   \includegraphics[width=0.98\textwidth]{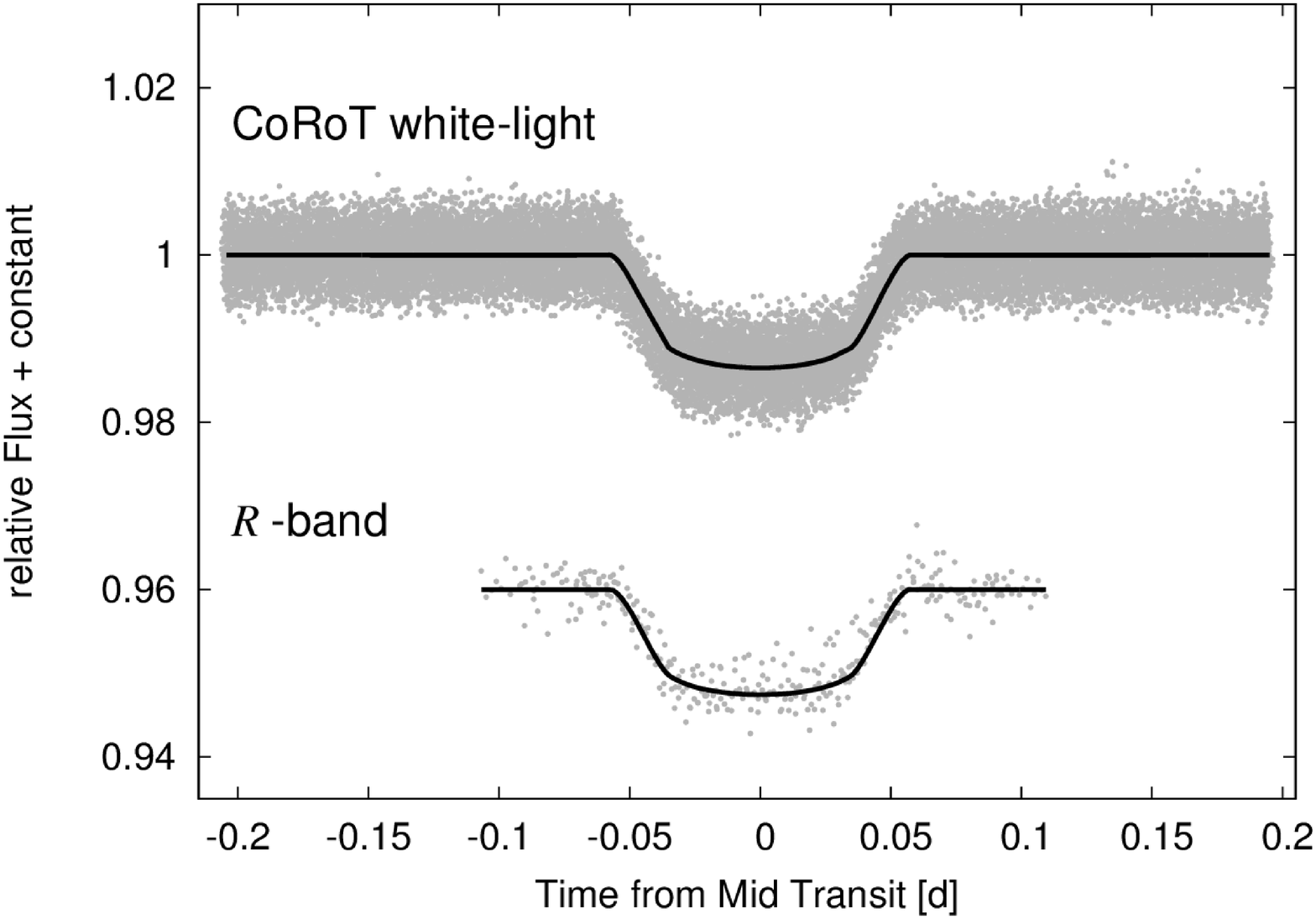}
  \caption{Phase-folded LCs of all 31 \textit{CoRoT} transits as well as of all three OSN $R$-band transits of CoRoT-5. The trend was removed before phase-folding. Overlaid are the best-fitting models obtained with TAP. }
  \label{alltransit_phased_C5}
\end{minipage}
\end{figure}

CoRoT-5\,b was discovered during the first LR on the galactic anti-centre direction (LRa01) that started on 2007 October 24 and lasted 112\,d \citep{2009A&A...506..281R}. The observations started with a cadence of 512s that was changed to the 32s-mode after seven transit events. In total, 31 transits were found. One event was lost in a larger data gap caused by a DPU reset \citep{2009A&A...506..281R}. Photometric and spectroscopic follow-up observations were scheduled right after the ``alarm mode'' was triggered. Radial velocity measurement with SOPHIE and HARPS confirmed that CoRoT-5\,b is a hot Jupiter-type planet orbiting a 14\,mag F9V star. The spectroscopic observations also yielded a slight eccentricity of the planetary orbit. The published physical properties of the CoRoT-5 system are summarized in Table~\ref{Werte_CoRoT}. CoRoT-5\,b belongs to the planets with the lowest mean density. It was found to be larger by 20\% compared to standard evolutionary models \citep{2009A&A...506..281R}. \\ We observed three ground-based LCs between 2014 January and 2016 December at OSN. The exposure times of the $R$-band observations were chosen between 120 and 150\,s. \\ We also re-analysed the \textit{CoRoT} data that initially consisted of 269\,390 data points. After removing all flagged data the number of data points reduced to 236\,774. We extracted 31 transit events with a depth of $\sim$1.4\% and a duration of $\sim$2.7\,h. The contribution of $L_{\mathrm{3}}$ was estimated to be 8.4\% by \citet{2009A&A...506..281R}.\\ The ground-based LCs together with the best-fitting model is shown in Fig.~\ref{LC_CoRoT5}. We have also plotted the transit times into an O--C diagram. Our ground-based observations deviate up to $\sim$20\,min from the predicted transit times, while the uncertainties on the original ephemeris estimated a shift of only $\sim$2.5\,min. Since an accurate determination of the ephemeris is hindered by the short time span of the \textit{CoRoT} observations, the uncertainties on the original ephemeris might have been underestimated. An alternative explanation for the deviation of one order of magnitude more than predicted could be the presence of significant transit timing variations (TTVs). This is, however, not supported by our observations. The result of our re-determined ephemeris is given in equation \ref{Elemente_CoRoT5}, where $E$ denotes the epoch ($\chi^{2}$\,=\,8.8, reduced $\chi^{2}$\,=\,0.27):
\begin{equation}
\label{Elemente_CoRoT5}
\begin{array}{r@{.}lcr@{.}l}
T_{\mathrm{c[BJD_{TDB}]}}(E)=(2454400 & 19896 & + & E\cdot 4 & 0379156)\,\mathrm{d} \\ 
\pm 0 & 00022 &  & \pm 0 & 0000012
\end{array}
\end{equation}
The updated version of the O--C diagram as well as all transit times and the corresponding O--C values are given in Fig.~\ref{O_C_CoRoT5} and Table~\ref{CoRoT5_Transit_Times}, respectively. We could not find any periodic signal in the transit times. The highest peak in the periodogram obtained with \begin{scriptsize}GLS\end{scriptsize} at a period of $P_{\mathrm{TTV}}$\,=\,90.0\,$\pm$\,0.8\,epochs shows a false-alarm-probability (FAP) of 99.9\%. The asymmetric shape of our LC from 2016 Dec. 20 could be an indication of stellar activity. Large spots are, however, unlikely as CoRoT-5 does not show strong out of transit variability \citep{2009A&A...506..281R}. We cannot discard that the CoRoT observations were taken during a minimum of the stellar activity cycle, and the ground-based observations are carried out during a maximum. Further high-precision photometric follow-up observations would be necessary to confirm stellar activity.

\begin{table}
\centering
\caption{Transit times for all observed transits of CoRoT-5\,b including the re-analysed \textit{CoRoT} transits. The O--C was calculated with the ephemeris given in equation \ref{Elemente_CoRoT5}.}
\label{CoRoT5_Transit_Times}
\renewcommand{\arraystretch}{1.1}
\begin{tabular}{ccr@{\,$\pm$\,}lr@{\,$\pm$\,}l}
\hline \hline
Telescope & Epoch & \multicolumn{2}{c}{$T_{\mathrm{c}}$ [BJD$_{\mathrm{TDB}}$]} & \multicolumn{2}{c}{O--C  [min]} \\ \hline \hline
CoRoT & 0	& 2454400.1999	& $^{0.0018}_{0.0018}$  & 1.42  & $^{2.65}_{2.54}$ \\
CoRoT & 1	& 2454404.2360	& $^{0.0015}_{0.0015}$  & -1.19 & $^{2.17}_{2.17}$ \\
CoRoT & 2	& 2454408.2756	& $^{0.0013}_{0.0013}$  & 1.17  & $^{1.83}_{1.86}$ \\
CoRoT & 3	& 2454412.3138	& $^{0.0015}_{0.0015}$  & 1.62  & $^{2.16}_{2.14}$ \\
CoRoT & 4	& 2454416.3486	& $^{0.0019}_{0.0020}$  & -2.85 & $^{2.75}_{2.85}$ \\
CoRoT & 5	& 2454420.3881	& $^{0.0014}_{0.0014}$  & -0.65 & $^{1.96}_{1.98}$ \\
CoRoT & 6	& 2454424.4250	& $^{0.0016}_{0.0016}$  & -2.08 & $^{2.33}_{2.24}$ \\
CoRoT & 7	& 2454428.4639	& $^{0.0009}_{0.0009}$  & -0.69 & $^{1.30}_{1.30}$ \\
CoRoT & 8	& 2454432.5027	& $^{0.0012}_{0.0012}$  & 0.63  & $^{1.70}_{1.69}$ \\
CoRoT & 9	& 2454436.5409	& $^{0.0013}_{0.0012}$  & 0.99  & $^{1.82}_{1.80}$ \\
CoRoT & 10	& 2454440.5783	& $^{0.0010}_{0.0010}$  & 0.22  & $^{1.51}_{1.45}$ \\
CoRoT & 11	& 2454444.6162	& $^{0.0010}_{0.0010}$  & 0.22  & $^{1.50}_{1.49}$ \\
CoRoT & 12	& 2454448.6538	& $^{0.0013}_{0.0012}$  & -0.14 & $^{1.80}_{1.74}$ \\
CoRoT & 13	& 2454452.6925	& $^{0.0010}_{0.0010}$  & 0.96  & $^{1.50}_{1.49}$ \\
CoRoT & 14	& 2454456.7289	& $^{0.0010}_{0.0010}$  & -1.25 & $^{1.40}_{1.45}$ \\
CoRoT & 15	& 2454460.7680	& $^{0.0011}_{0.0011}$  & 0.45  & $^{1.60}_{1.56}$ \\
CoRoT & 16	& 2454464.8056	& $^{0.0011}_{0.0011}$  & 0.01  & $^{1.60}_{1.61}$ \\
CoRoT & 17	& 2454468.8431	& $^{0.0012}_{0.0012}$  & -0.56 & $^{1.78}_{1.78}$ \\
CoRoT & 18	& 2454472.8810	& $^{0.0011}_{0.0013}$  & -0.64 & $^{1.65}_{1.82}$ \\
CoRoT & 19	& 2454476.9187	& $^{0.0010}_{0.0010}$  & -0.94 & $^{1.50}_{1.50}$ \\
CoRoT & 20	& 2454480.9570	& $^{0.0011}_{0.0012}$  & -0.42 & $^{1.65}_{1.66}$ \\
CoRoT & 22	& 2454489.0327	& $^{0.0011}_{0.0010}$  & -0.53 & $^{1.53}_{1.50}$ \\
CoRoT & 23	& 2454493.0712	& $^{0.0010}_{0.0010}$  & 0.21  & $^{1.42}_{1.48}$ \\
CoRoT & 24	& 2454497.1089	& $^{0.0028}_{0.0029}$  & 0.02  & $^{4.10}_{4.13}$ \\
CoRoT & 25	& 2454501.1465	& $^{0.0012}_{0.0012}$  & -0.57 & $^{1.67}_{1.67}$ \\
CoRoT & 26	& 2454505.1854	& $^{0.0012}_{0.0012}$  & 0.85  & $^{1.71}_{1.70}$ \\
CoRoT & 27	& 2454509.2233	& $^{0.0011}_{0.0011}$  & 0.87  & $^{1.59}_{1.58}$ \\
CoRoT & 28	& 2454513.2618	& $^{0.0011}_{0.0011}$  & 1.74  & $^{1.63}_{1.57}$ \\
CoRoT & 29	& 2454517.2987	& $^{0.0011}_{0.0012}$  & 0.30  & $^{1.60}_{1.69}$ \\
CoRoT & 30	& 2454521.3361	& $^{0.0011}_{0.0011}$  & -0.49 & $^{1.57}_{1.59}$ \\
CoRoT & 31	& 2454525.3752	& $^{0.0023}_{0.0023}$  & 1.27  & $^{3.36}_{3.31}$ \\
OSN   & 561	& 2456665.4696	& $^{0.0007}_{0.0006}$  & 0.00  & $^{0.99}_{0.91}$ \\
OSN   & 724	& 2457323.6502	& $^{0.0024}_{0.0025}$  & 0.57  & $^{3.44}_{3.53}$ \\
OSN   & 828	& 2457743.5912	& $^{0.0049}_{0.0052}$  & -2.63 & $^{6.99}_{7.52}$ \\
\hline \hline
\end{tabular}
\end{table}

\begin{figure}
  \centering
  \includegraphics[width=0.3\textwidth,angle=270]{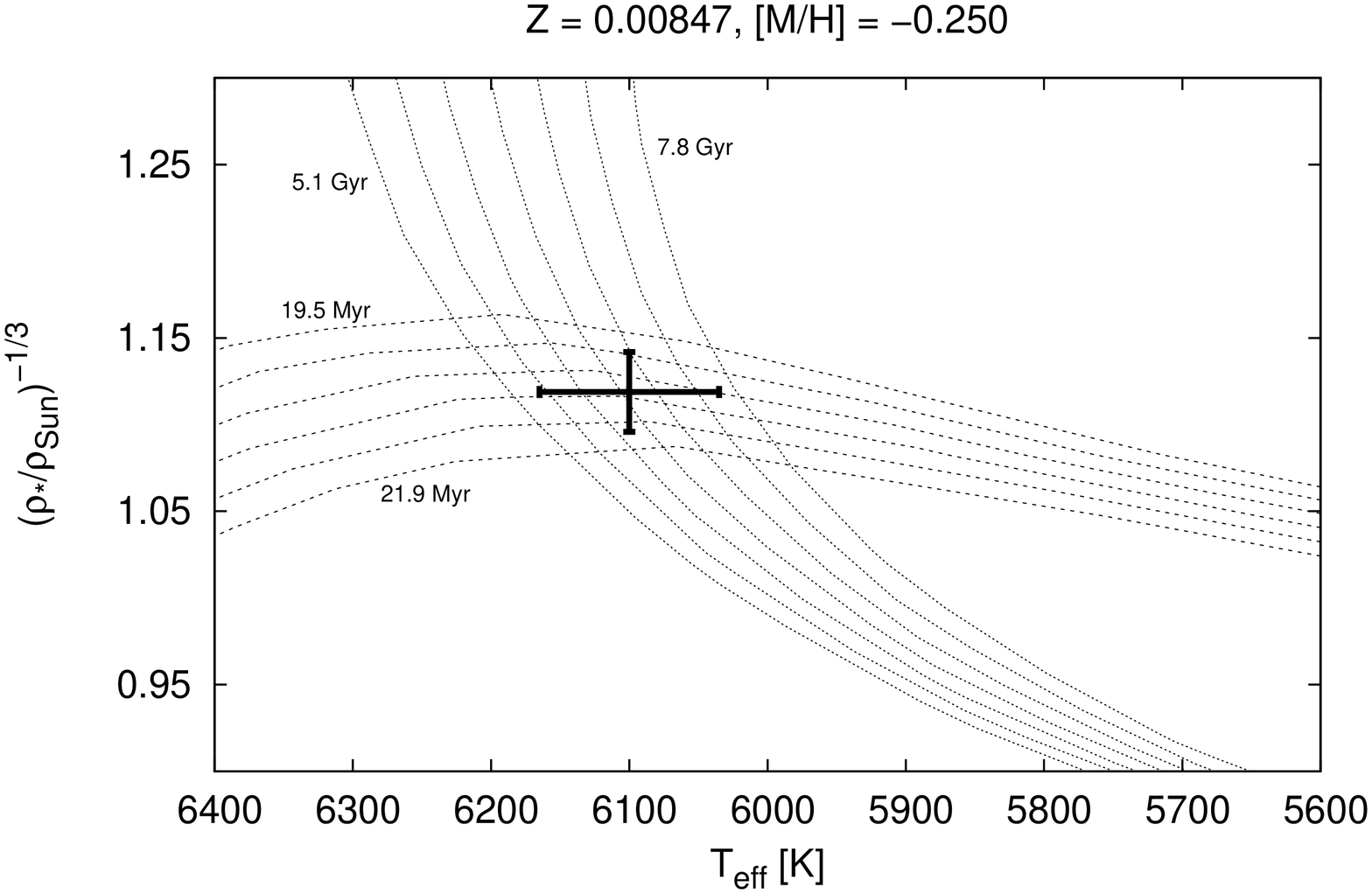}
  \caption{Position of CoRoT-5 in the $\rho_{\mathrm{A}}^{-1/3}\,-\,T_{\mathrm{eff}}$ plane. The PARSEC isochrones of metallicity [M/H]=\,-0.25 for log(age)\,=\,7.29\,-\,7.34 with steps of 0.01 and log(age)\,=\,9.71\,-\,9.89 with steps of 0.03 for the young and the old age, respectively, are also shown.}
  \label{HRD_C5}
\end{figure}

\noindent The results of the LC analysis is given in Table~\ref{tbl:TAPmcmc1} and shown in Fig. \ref{alltransit_phased_C5}, the obtained physical properties are summarized in Table~\ref{phys_prop_CoRoT5}. We found the geometrical parameters in excellent agreement with the ones of \citet{2009A&A...506..281R}. Also most of the stellar and planetary values agree with each other within their error bars on a 2$\sigma$ level. Only the surface gravity of the star that was determined spectroscopically in \citet{2009A&A...506..281R}  slightly differs. These authors also give the photometrically obtained value of log\,$g_{\mathrm{A}}$=4.311$\pm$0.033, which agrees with the result of our LC analysis. \\ Fig.~\ref{HRD_C5} shows the position of CoRoT-5 in a modified version of the HRD, together with the PARSEC isochrones. CoRoT-5 is in an area of the HRD with overlapping isochrones of young ($\sim$\,20\,Myr) and old ($\sim$\,6\,Gyr) ages. However, as \citet{2009A&A...506..281R} have already shown, the very low level of stellar variability in the global LC as well as the missing signs of the CaII or a strong LiI absorption line hints to the older age.


\begin{table}
\centering
\caption{Physical properties of the CoRoT-5 system derived from LC modelling. Values derived by \citet[][R09]{2009A&A...506..281R} and \citet[][S11]{2011MNRAS.417.2166S} are given for comparison.}
\label{phys_prop_CoRoT5}
\renewcommand{\arraystretch}{1.1}
\begin{tabular}{lr@{\,$\pm$\,}lr@{\,$\pm$\,}lr@{\,$\pm$\,}l}
\hline \hline
 Parameter & \multicolumn{2}{c}{This work} & \multicolumn{2}{c}{R09} & \multicolumn{2}{c}{S11} \\ \hline \hline
& \multicolumn{6}{c}{Planetary parameters} \\ \hline 
$R_{\mathrm{b}}$  [R$_{\mathrm{Jup}}$] & 1.256 & $^{0.046}_{0.045}$ & 1.388 & $^{0.046}_{0.047}$ & 1.182 & $^{0.102}_{0.098}$ \\
$M_{\mathrm{b}}$  [M$_{\mathrm{Jup}}$] & 0.459 & $^{0.053}_{0.032}$ & 0.467 & $^{0.047}_{0.024}$ & 0.470 & $^{0.058}_{0.031}$ \\
$\rho_{\mathrm{b}}$  [$\mathrm{\rho}_{\mathrm{Jup}}$] & 0.217 & $^{0.035}_{0.028}$ & 0.163 & $^{0.023}_{0.019}$ & 0.266 & $^{0.082}_{0.058}$ \\
log\,$g_{\mathrm{b}}$ & 2.86 & $^{0.05}_{0.03}$ & 2.89 & $^{0.08}_{0.05}$  & 2.92 & $^{0.09}_{0.07}$ \\
$T_{\mathrm{eq}}$ [K] & 1397 & $^{15}_{15}$ & 1438 & 38 & 1348 & $^{50}_{51}$ \\ 
$\Theta$ & 0.0366 & $^{0.0051}_{0.0038}$ & \multicolumn{2}{c}{} & 0.0388 & $^{0.0054}_{0.0038}$ \\ \hline
& \multicolumn{6}{c}{Stellar parameters} \\ \hline 
$R_{\mathrm{A}}$  [R$_{\mathrm{\odot}}$] & 1.115 & $^{0.035}_{0.033}$ & 1.186 & 0.04 & 1.052 & $^{0.081}_{0.067}$ \\
$M_{\mathrm{A}}$  [M$_{\mathrm{\odot}}$] & 0.99 & 0.07 & 1.00 & 0.02 & 1.025 & $^{0.100}_{0.056}$ \\
$\rho_{\mathrm{A}}$  [$\mathrm{\rho}_{\mathrm{\odot}}$] & 0.714 & $^{0.045}_{0.043}$ & \multicolumn{2}{c}{} & 0.88 & $^{0.21}_{0.16}$ \\ 
log\,$g_{\mathrm{A}}$ & 4.339 & $^{0.021}_{0.020}$ & 4.19 & 0.03 & 4.405 & $^{0.068}_{0.059}$ \\
log$\frac{L_{\mathrm{A}}}{L_{\mathrm{\odot}}}$ & 0.18 & 0.05 & \multicolumn{2}{c}{} & \multicolumn{2}{c}{} \\
log(Age) & 7.31 & 0.03 & \multicolumn{2}{c}{9.74 - 9.92} & \multicolumn{2}{c}{} \\
  & 9.80 & 0.12 & \multicolumn{2}{c}{} & \multicolumn{2}{c}{} \\\hline
& \multicolumn{6}{c}{Geometrical parameters} \\ \hline 
$a$  [au] & \multicolumn{2}{c}{0.0495}  & \multicolumn{2}{c}{0.04947}  & \multicolumn{2}{c}{0.0500}  \\
& & 0.0012 & &  $^{0.00026}_{0.00029}$ & & $^{0.0016}_{0.0009}$ \\
$i$  [$^{\circ}$] & 85.68 & $^{0.18}_{0.17}$ & 85.83 & $^{0.99}_{1.38}$ & 86.24 & 0.53 \\
$b$ & 0.76 & $^{0.05}_{0.04}$ & 0.755 & $^{0.017}_{0.022}$ & \multicolumn{2}{c}{} \\ \hline \hline
\end{tabular}
\end{table}

\section{CoRoT-8}

\begin{figure*}
  \includegraphics[width=0.235\textwidth, angle=270]{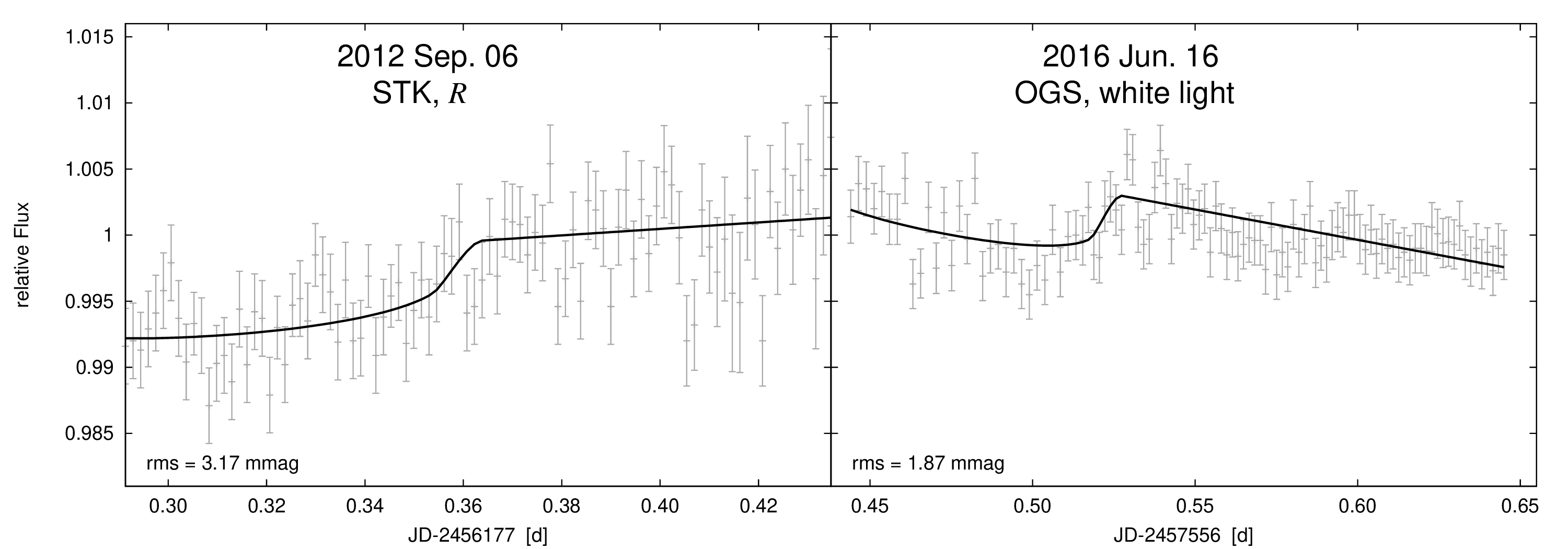}
  \caption{Same as Fig.~\ref{LC_CoRoT5} but for CoRoT-8. The date of observation, observatory, filter, and the $rms$ of the fit are indicated in each individual panel.}
\label{LC_CoRoT8}
\end{figure*}

\begin{figure}
\begin{minipage}[]{0.45\textwidth}
  \includegraphics[width=0.98\textwidth]{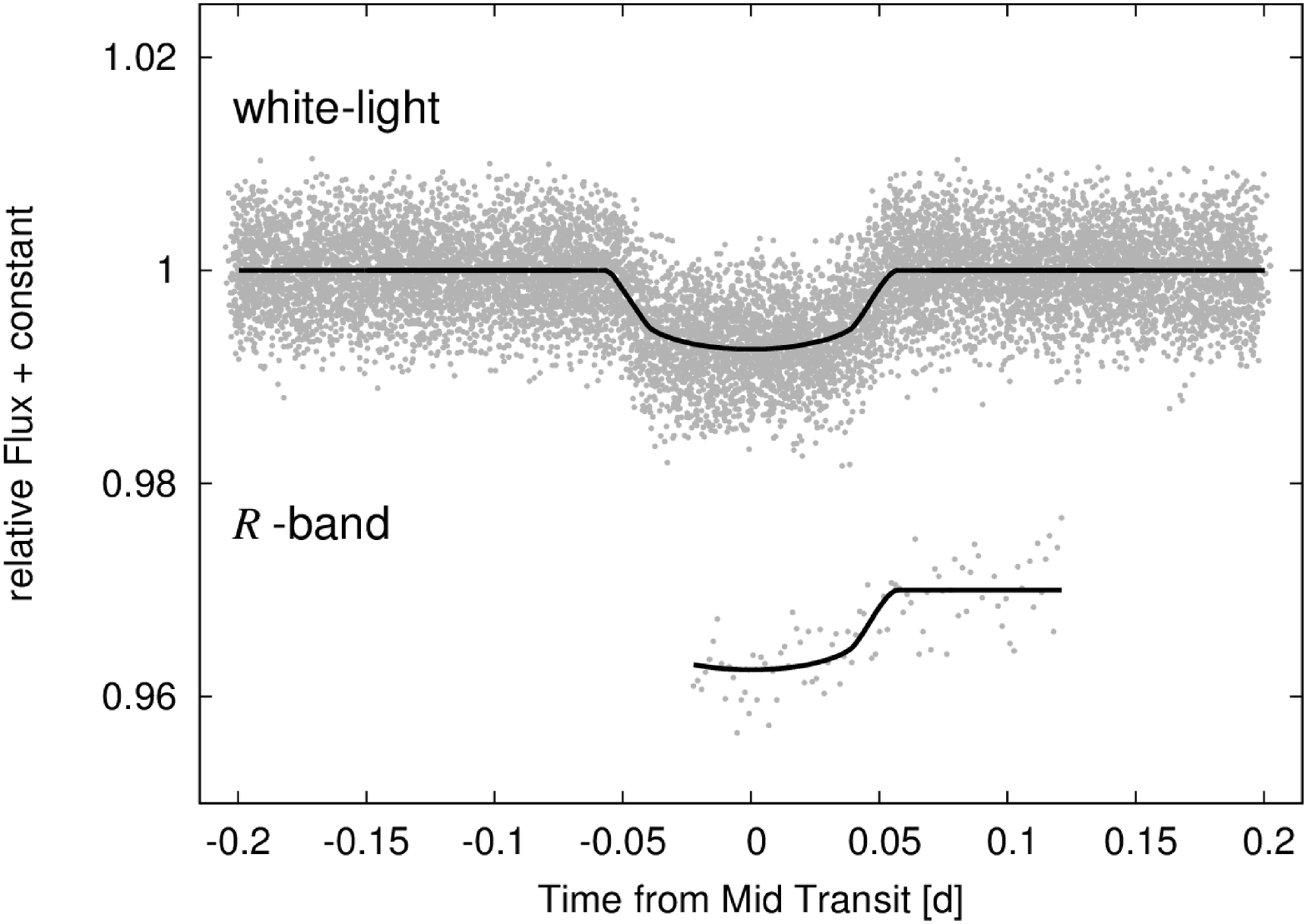}
  \caption{Phase-folded LCs of all 23 \textit{CoRoT} transits as well as our own transits of CoRoT-8. The trend was removed before phase-folding. Overlaid are the best-fitting models obtained with TAP. }
  \label{alltransit_phased_C8} 
\end{minipage}
\begin{minipage}[]{0.45\textwidth}
   \includegraphics[width=0.65\textwidth,angle=270]{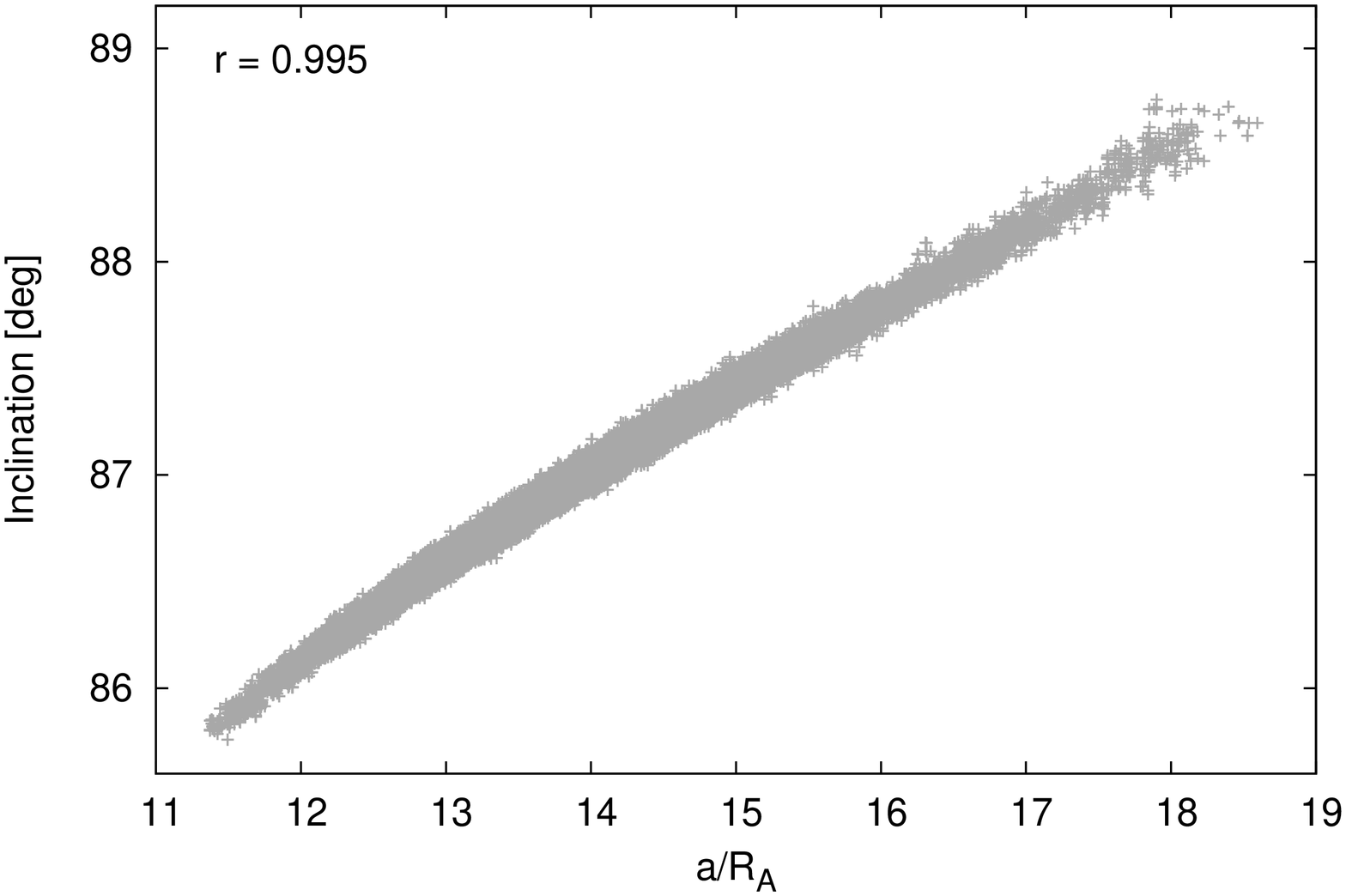}
  \caption{Example of a significant correlation between $a$/$R_{\mathrm{A}}$ and the orbital inclination $i$ (correlation coefficients $r=0.995$) of CoRoT-8\,b for one MCMC chain.}
  \label{Corr_C8}
\end{minipage}

\end{figure}

CoRoT-8b, which was observed by \textit{CoRoT} during the first LR in constellation Aquila (LRc01) from 2007 May 16 to October 5, orbits a K1 dwarf in $\sim$6.2\,d \citep{2010A&A...520A..66B}. It was detected by the `alarm mode'-pipeline which switched the observation mode to the short cadence after $\sim$68\,d and triggered follow-up observations. $L_{\mathrm{3}}$ is given as 1.55\% in the \textit{Exo-dat} data base \citep{2009AJ....138..649D}. RV follow-up observations that confirmed the planetary nature of CoRoT-8\,b were carried out with SOPHIE and HARPS. With its measured radius and mass, CoRoT-8\,b appears to be somewhat between Saturn and Neptune \citep{2010A&A...520A..66B}. \\ We observed two transits of CoRoT-8 separated by $\sim$4\,yr. One transit was observed in 2012 September 6 at the University Observatory Jena and the other one in 2016 June 16 at ESA's OGS. Unfortunately, in both cases we could only observe a partial transit event. Our LCs are shown in Fig.~\ref{LC_CoRoT8}.\\ We extracted 23 transits, 11 of them in long cadence mode, from the \textit{CoRoT}-LC that consist in total of 182\,380 unflagged data points. The altogether 25 transits were simultaneously modelled. The phase-folded LCs including all transits are shown in Fig.~\ref{alltransit_phased_C8}. The resulting system parameters are given in Table~\ref{tbl:TAPmcmc1}.\\ Computing the physical properties of the system from these system parameters resulted in significant deviations from the values given in \citet{2010A&A...520A..66B}. In particular, the stellar radius $R_{\mathrm{A}}=$1.048$\pm^{0.082}_{0.067}$ R$_{\mathrm{\odot}}$, the stellar density  $\rho_{\mathrm{A}}=$0.89$\pm^{0.20}_{0.15}$ $\mathrm{\rho}_{\mathrm{\odot}}$ and the impact parameter $b=$0.75$\pm^{0.11}_{0.09}$ differ by more than 3-$\sigma$. The much lower density results in a higher stellar mass and a very low pre-main-sequence age of log(age)\,=\,7.38$\pm$0.13 when plotting it into the modified HRD with the PARSEC isochrones. Consequently, also the planetary parameters deviate. The discrepancies originate in the best-fitting values of $i$, $a$/$R_{\mathrm{A}}$ and $R_{\mathrm{b}}$/$R_{\mathrm{A}}$ obtained with \begin{scriptsize}TAP\end{scriptsize}, which we found to be strongly correlated. The analysis of relations between the parameters reveals significant correlation or anti-correlations (with the Pearson correlation coefficients $r$ ranging from 0.873 to 0.995) between $i$ and $a$/$R_{\mathrm{A}}$, $i$ and $R_{\mathrm{b}}$/$R_{\mathrm{A}}$, and $a$/$R_{\mathrm{A}}$ and $R_{\mathrm{b}}$/$R_{\mathrm{A}}$. An example for the correlation between  $i$ and $a$/$R_{\mathrm{A}}$ is shown in Fig.~\ref{Corr_C8}. \\ CoRoT-8 was found to be a  K1 main-sequence star by \citep{2010A&A...520A..66B}. They excluded very young ages because of its slow rotation and the absence of detectable CaII or LiI absorption lines. Hence, a radius of $\sim$1\,R$_{\mathrm{\odot}}$ is very unlikely. A cross-check with Gaia DR2 yielded a radius of 0.8 (0.71\,-\,0.87) R$_{\mathrm{\odot}}$ \citep[][note: this value has to be taken with caution]{2016A&A...595A...1G,2018A&A...616A...1G}. Because of the strong parameter correlations a smaller radius (a higher $a$/$R_{\mathrm{A}}$) can be accounted for with a higher inclination $i$ without degrading the quality of the fit. Therefore, we placed a prior before we re-fit our data. \citet{2010A&A...520A..66B} determined a projected stellar rotational velocity of $v$\,sin\,$i=2\pm1$\,km\,s$^{-1}$. Using the gyrochronology relation by \citet{2015MNRAS.450.1787A} and assuming spin-orbit alignment ($i\sim90^{\circ}$), we estimated a stellar age of 1.7$\pm^{2.3}_{1.4}$\,Gyr. The PARSEC isochrones of our age estimate were used to constrain the stellar density to $\rho_{\mathrm{A}}=\,1.73\,\pm\,0.26\,\mathrm{\rho}_{\mathrm{\odot}}$. The resulting value of $a$/$R_{\mathrm{A}}=17.07\pm0.84$ \citep[calculated by using the formula of][]{2010arXiv1001.2010W} was finally used as prior for the LC modelling with \begin{scriptsize}TAP\end{scriptsize}. The results of our re-analysis using a prior on the stellar density are given in Table~\ref{CoRoT8_TAP_neu}, and the corresponding physical properties of the system in comparison to the literature values are summarized in Table~\ref{phys_prop_CoRoT8}. By applying gyrochronology to constrain the stellar density we found the physical properties in good agreement (on a 2-$\sigma$ level) with the values of \citet{2010A&A...520A..66B}. However, using a prior in the fitting process may result in our uncertainties being underestimated.

\begin{table}
\centering
\caption{System parameters for CoRoT-8 resulting from the LC analysis with TAP. Unlike the values in Table~\ref{tbl:TAPmcmc1}, $a$/$R_{\mathrm{A}}$ was only allowed to vary around the value derived from a prior on the stellar density, under the Gaussian penalty defined by the derived error.}
\label{CoRoT8_TAP_neu}
\begin{tabular}{lr@{\,$\pm$\,}l}
\hline \hline
Parameter & \multicolumn{2}{c}{Value} \\ \hline
Inclination [$^{\circ}$] & 88.178 & $^{0.083}_{0.082}$\\
$a$/$R_{\mathrm{A}}$ $^{a}$ & 17.05 & $^{0.16}_{0.17}$\\
$R_{\mathrm{b}}$/$R_{\mathrm{A}}$ (\textit{CoRoT} white light) & 0.07915 & $^{0.00099}_{0.00098}$\\
$R_{\mathrm{b}}$/$R_{\mathrm{A}}$ (\textit{R}-band) & 0.072 & $^{0.0120}_{0.0081}$\\
$R_{\mathrm{b}}$/$R_{\mathrm{A}}$ (white light) & 0.0691 & $^{0.0093}_{0.0064}$\\
Linear LD$^{b}$ (\textit{CoRoT} white light) & 0.583 & $^{0.021}_{0.021}$\\
Quad LD$^{b}$ (\textit{CoRoT} white light) & 0.133 & $^{0.021}_{0.020}$\\
Linear LD$^{b}$ (\textit{R}-band) & 0.498  & $^{0.094}_{0.098}$\\
Quad LD$^{b}$ (\textit{R}-band) & 0.226 & $^{0.097}_{0.098}$\\
Linear LD$^{b}$ (white light) & 0.36 & $^{0.100}_{0.098}$\\
Quad LD$^{b}$ (white light) & 0.271 & $^{0.098}_{0.099}$\\  \hline \hline 
\end{tabular}
\\$^{a}$ was allowed to vary within a prior
\\$^{b}$ were allowed to vary $\pm0.1$ around the theoretical values
\end{table}

\begin{table}
\centering
\caption{Physical properties of the CoRoT-8 system derived from the results of the LC modelling given in Table~\ref{CoRoT8_TAP_neu} based on constraints on the stellar density. Values derived by \citet[][B10]{2010A&A...520A..66B} and \citet[][S11]{2011MNRAS.417.2166S} are given for comparison.}
\label{phys_prop_CoRoT8}
\renewcommand{\arraystretch}{1.1}
\begin{tabular}{lr@{\,$\pm$\,}lr@{\,$\pm$\,}lr@{\,$\pm$\,}l}
\hline \hline
 Parameter & \multicolumn{2}{c}{This work} & \multicolumn{2}{c}{B10} & \multicolumn{2}{c}{S11} \\ \hline \hline
& \multicolumn{6}{c}{Planetary parameters} \\ \hline 
$R_{\mathrm{b}}$  [R$_{\mathrm{Jup}}$] 	&	0.619	&	 $^{0.016}_{0.017}$ 	&	0.57	&	0.02	&	0.712	&	0.083 \\
$M_{\mathrm{b}}$  [M$_{\mathrm{Jup}}$] 	&	0.218	&	 $^{0.034}_{0.034}$ 	&	0.22	&	0.03	&	0.216	&	0.036 \\
$\rho_{\mathrm{b}}$  [$\mathrm{\rho}_{\mathrm{Jup}}$] 	&	0.86	&	 $^{0.15}_{0.15}$ 	&	1.20	&	0.08	&	0.56	&	0.21 \\
log\,$g_{\mathrm{b}}$ 	&	3.15	&	 $^{0.07}_{0.07}$ 	&	 \multicolumn{2}{c}{} 			&	3.03	&	0.12 \\
$T_{\mathrm{eq}}$ [K] 	&	870	&	 $^{14}_{14}$ 	&	 \multicolumn{2}{c}{} 			&	922	&	41 \\
$\Theta$ 	&	0.0503	&	 $^{0.0083}_{0.0083}$ 	&	 \multicolumn{2}{c}{} 			&	0.0437	&	0.0084 \\ \hline
	&	 \multicolumn{6}{c}{Stellar parameters} \\ \hline 										
$R_{\mathrm{A}}$  [R$_{\mathrm{\odot}}$] 	&	0.802	&	 $^{0.014}_{0.014}$ 	&	0.77	&	0.02	&	0.898	&	0.090 \\
$M_{\mathrm{A}}$  [M$_{\mathrm{\odot}}$] 	&	0.89	&	0.04	&	0.88	&	0.04	&	0.878	&	0.078 \\
$\rho_{\mathrm{A}}$  [$\mathrm{\rho}_{\mathrm{\odot}}$] 	&	1.73	&	 $^{0.26}_{0.26}$ 	&	1.91	&	0.07	&	1.21	&	0.32 \\
log\,$g_{\mathrm{A}}$ 	&	4.58	&	 $^{0.01}_{0.01}$ 	&	4.58	&	0.08	&	4.475	&	0.077 \\
log$\frac{L_{\mathrm{A}}}{L_{\mathrm{\odot}}}$ 	&	-0.40	&	0.05	&	 \multicolumn{2}{c}{} 			&	 \multicolumn{2}{c}{} \\		
log(Age) 	&	9.23	&	$^{0.37}_{0.75}$*	&	 \multicolumn{2}{c}{$\leq$9.48} 			&	 \multicolumn{2}{c}{unconstrained} \\	\hline
	&	 \multicolumn{6}{c}{Geometrical parameters} \\ \hline 										
$a$  [au] 	&	 \multicolumn{2}{c}{0.0636}  			&	 \multicolumn{2}{c}{0.063}  			&	 \multicolumn{2}{c}{0.0633}  \\		
	&	 	&	0.0014	&	 	&	0.001	&	 	&	0.0019 \\
$i$  [$^{\circ}$] 	&	88.18	&	 $^{0.08}_{0.08}$ 	&	88.4	&	0.1	&	87.44	&	 0.56 \\
$b$ 	&	0.54	&	 $^{0.03}_{0.02}$ 	&	0.49	&	0.04	&	 \multicolumn{2}{c}{} \\ \hline \hline		
\end{tabular}
\\$^{\ast}$ determined by gyrochronology
\end{table}

\begin{figure}
  \centering
  \includegraphics[width=0.28\textwidth,angle=270]{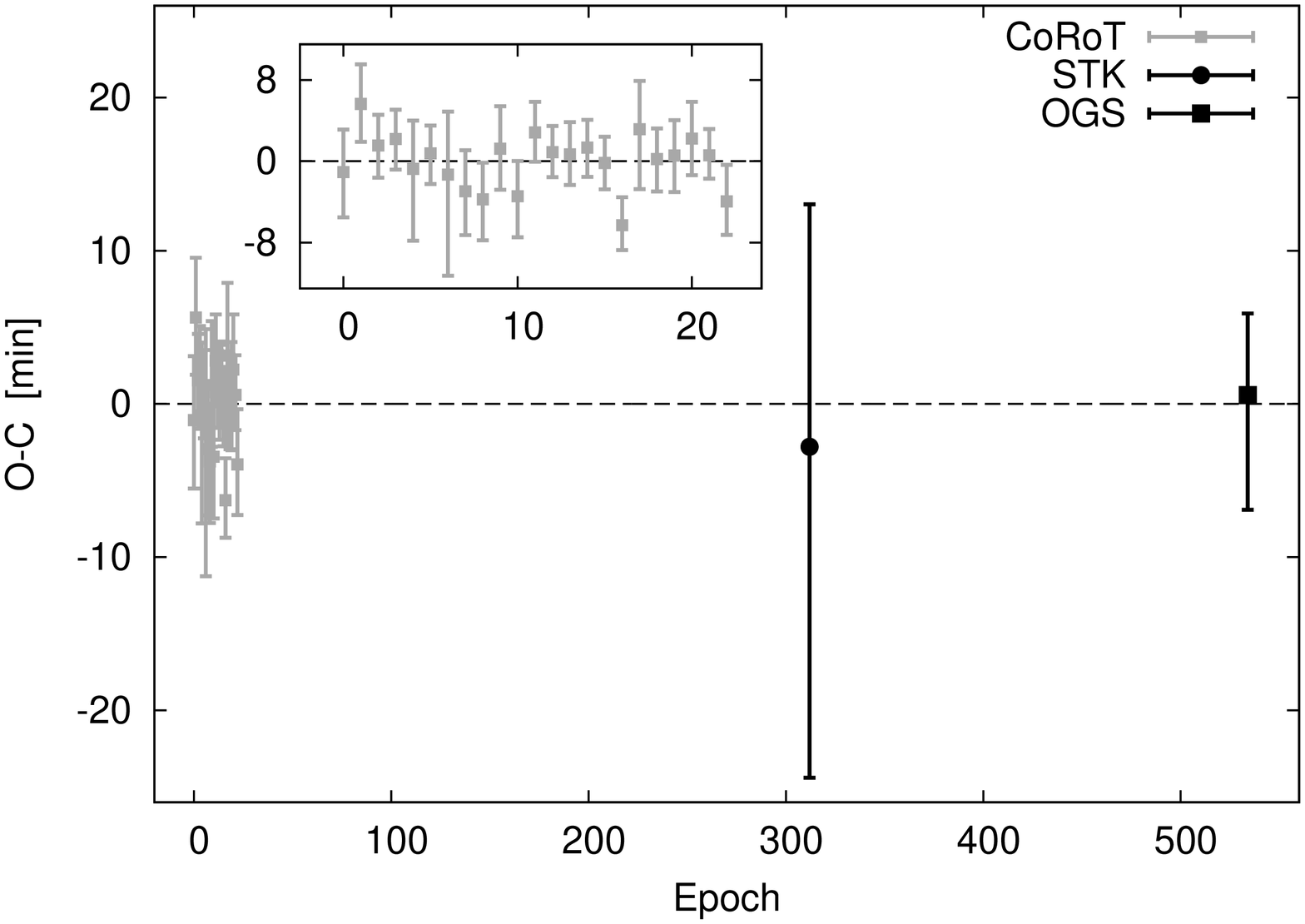}
  \caption{Same as Fig,~\ref{O_C_CoRoT5}  but for CoRoT-8\,b. The dashed line represents the updated ephemeris given in equation \ref{Elemente_CoRoT8}. }
  \label{O_C_CoRoT8}
\end{figure}

\noindent Although with large uncertainties in the transit times because of the partial transit coverage, our measurements deviate by up to 49\,min in the O--C diagram from the ephemeris given in \citet{2011MNRAS.417.2166S}, which is larger than the estimated uncertainty (see Fig. \ref{period_err}). Hence, we re-determined the ephemeris. The result is given in equation \ref{Elemente_CoRoT8} ($\chi^{2}$\,=\,14.6, reduced $\chi^{2}$\,=\,0.63):
\begin{equation}
\label{Elemente_CoRoT8}
\begin{array}{r@{.}lcr@{.}l}
T_{\mathrm{c[BJD_{TDB}]}}(E)=(2454239 & 03317 & + & E\cdot 6 & 212445)\,\mathrm{d} \\ 
\pm 0 & 00049 &  & \pm 0 & 000007
\end{array}
\end{equation}
The updated O--C diagram is shown in Fig.~\ref{O_C_CoRoT8}, and all transit times and O--C values are given in Table~\ref{CoRoT8_Transit_Times}. Our measurements are in very good agreement with the refined ephemeris. \citet{2010A&A...520A..66B} detected statistically significant TTVs within the \textit{CoRoT}-LC with a period of $\sim$7 Epochs ($\sim$43.5\,d). They claimed that, since it is close to a multiple of the stellar rotation period of $\sim$20\,d, the TTVs are induced by the stellar activity. With our analysis we cannot confirm these variations. Our period search in the O--C values with \begin{scriptsize}GLS\end{scriptsize} showed no significant signal. The highest peak with a period of $P_{\mathrm{TTV}}$\,=\,101.3\,$\pm$\,1.5\,epochs shows a FAP of 99.99\%.

\begin{table}
\centering
\caption{Same as Table~\ref{CoRoT5_Transit_Times} but for all transits of CoRoT-8\,b. The O--C was calculated with the ephemeris given in equation \ref{Elemente_CoRoT8}.}
\label{CoRoT8_Transit_Times}
\renewcommand{\arraystretch}{1.1}
\begin{tabular}{ccr@{\,$\pm$\,}lr@{\,$\pm$\,}l}
\hline \hline
Telescope & Epoch & \multicolumn{2}{c}{$T_{\mathrm{c}}$ [BJD$_{\mathrm{TDB}}$]} & \multicolumn{2}{c}{O--C  [min]} \\ \hline \hline
CoRoT & 0	& 2454239.0324	& $^{0.0029}_{0.0031}$  & -1.06 & $^{4.18}_{4.46}$ \\
CoRoT & 1	& 2454245.2495	& $^{0.0027}_{0.0026}$  &  5.65 & $^{3.89}_{3.74}$ \\
CoRoT & 2	& 2454251.4591	& $^{0.0021}_{0.0022}$  &  1.55 & $^{3.02}_{3.17}$ \\
CoRoT & 3	& 2454257.6720	& $^{0.0020}_{0.0021}$  &  2.20 & $^{2.88}_{3.02}$ \\
CoRoT & 4	& 2454263.8824	& $^{0.0033}_{0.0049}$  & -0.74 & $^{4.75}_{7.06}$ \\
CoRoT & 5	& 2454270.0959	& $^{0.0019}_{0.0021}$  &  0.78 & $^{2.74}_{3.02}$ \\
CoRoT & 6	& 2454276.3069	& $^{0.0043}_{0.0069}$  & -1.31 & $^{6.19}_{9.94}$ \\
CoRoT & 7	& 2454282.5182	& $^{0.0028}_{0.0030}$  & -2.95 & $^{4.03}_{4.32}$ \\
CoRoT & 8	& 2454288.7301	& $^{0.0025}_{0.0028}$  & -3.74 & $^{3.60}_{4.03}$ \\
CoRoT & 9	& 2454294.9460	& $^{0.0029}_{0.0028}$  &  1.23 & $^{4.18}_{4.03}$ \\
CoRoT & 10	& 2454301.1552	& $^{0.0024}_{0.0028}$  & -3.44 & $^{3.46}_{4.03}$ \\
CoRoT & 11	& 2454307.3720	& $^{0.0021}_{0.0020}$  &  2.83 & $^{3.02}_{2.88}$ \\
CoRoT & 12	& 2454313.5831	& $^{0.0018}_{0.0017}$  &  0.89 & $^{2.59}_{2.45}$ \\
CoRoT & 13	& 2454319.7954	& $^{0.0022}_{0.0021}$  &  0.69 & $^{3.17}_{3.02}$ \\
CoRoT & 14	& 2454326.0083	& $^{0.0019}_{0.0020}$  &  1.34 & $^{2.74}_{2.88}$ \\
CoRoT & 15	& 2454332.2197	& $^{0.0018}_{0.0018}$  & -0.17 & $^{2.59}_{2.59}$ \\
CoRoT & 16	& 2454338.4279	& $^{0.0019}_{0.0017}$  & -6.28 & $^{2.74}_{2.45}$ \\
CoRoT & 17	& 2454344.6469	& $^{0.0033}_{0.0041}$  &  3.16 & $^{4.75}_{5.90}$ \\
CoRoT & 18	& 2454350.8573	& $^{0.0021}_{0.0022}$  &  0.21 & $^{3.02}_{3.17}$ \\
CoRoT & 19	& 2454357.0700	& $^{0.0024}_{0.0025}$  &  0.58 & $^{3.46}_{3.60}$ \\
CoRoT & 20	& 2454363.2836	& $^{0.0025}_{0.0025}$  &  2.24 & $^{3.60}_{3.60}$ \\
CoRoT & 21	& 2454369.4949	& $^{0.0018}_{0.0016}$  &  0.59 & $^{2.59}_{2.30}$ \\
CoRoT & 22	& 2454375.7042	& $^{0.0025}_{0.0023}$  & -3.94 & $^{3.60}_{3.31}$ \\
STK   & 312	& 2456177.3142	& $^{0.0110}_{0.0150}$  & -2.80 & $^{15.84}_{21.60}$ \\
OGS   & 534	& 2457556.4794	& $^{0.0037}_{0.0052}$  &  0.58 & $^{5.33}_{7.49}$ \\
\hline \hline
\end{tabular}
\end{table}

\section{CoRoT-12}

\begin{figure*}
  \includegraphics[width=0.235\textwidth, angle=270]{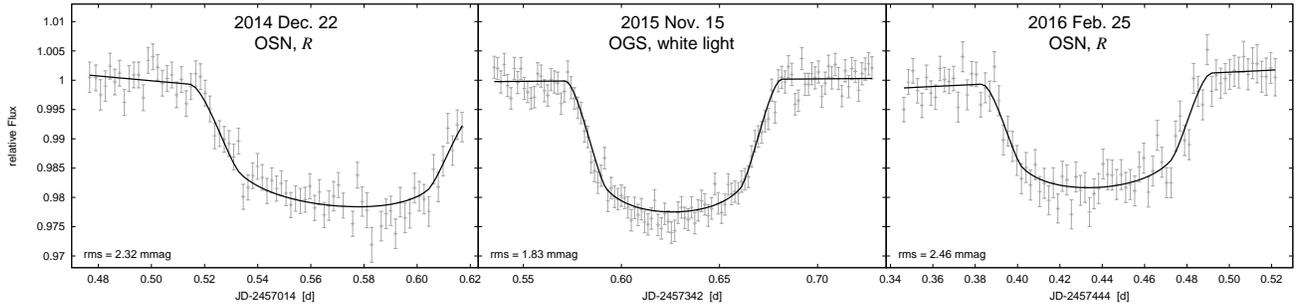}
  \caption{Same as Fig.~\ref{LC_CoRoT5} but for CoRoT-12. The date of observation, observatory, filter, and the $rms$ of the fit are indicated in each individual panel.}
\label{LC_CoRoT12}
\end{figure*}

\begin{figure}
\begin{minipage}[]{0.45\textwidth}
  \includegraphics[width=0.98\textwidth]{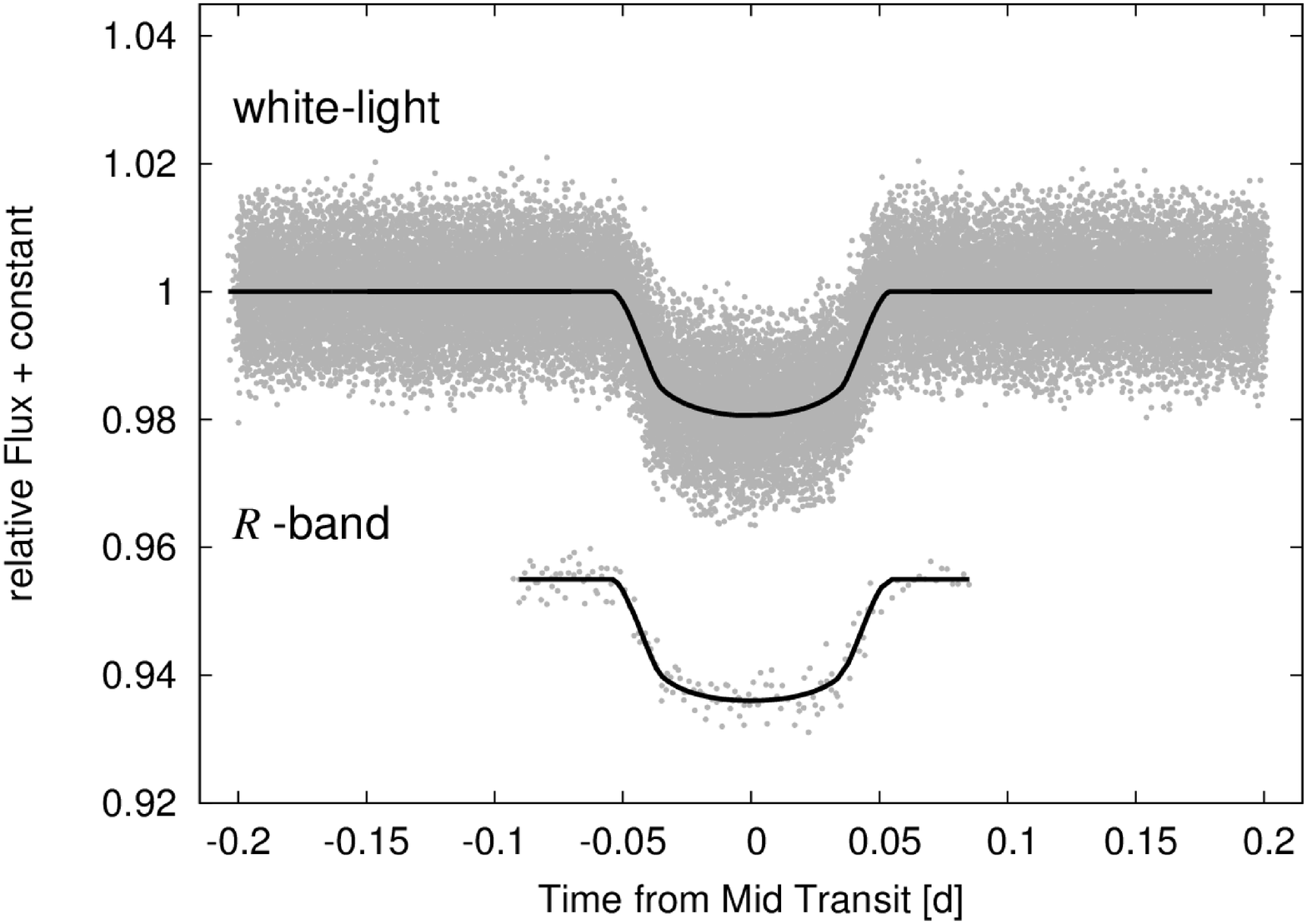}
  \caption{Phase-folded LCs of all 47 \textit{CoRoT} transits as well as our own transits of CoRoT-12. The trend was removed before phase-folding. Overlaid are the best-fitting models obtained with TAP.}
  \label{alltransit_phased_C12}
\end{minipage} 
\end{figure}

\begin{figure}
\begin{minipage}[]{0.45\textwidth}
  \includegraphics[width=0.65\textwidth,angle=270]{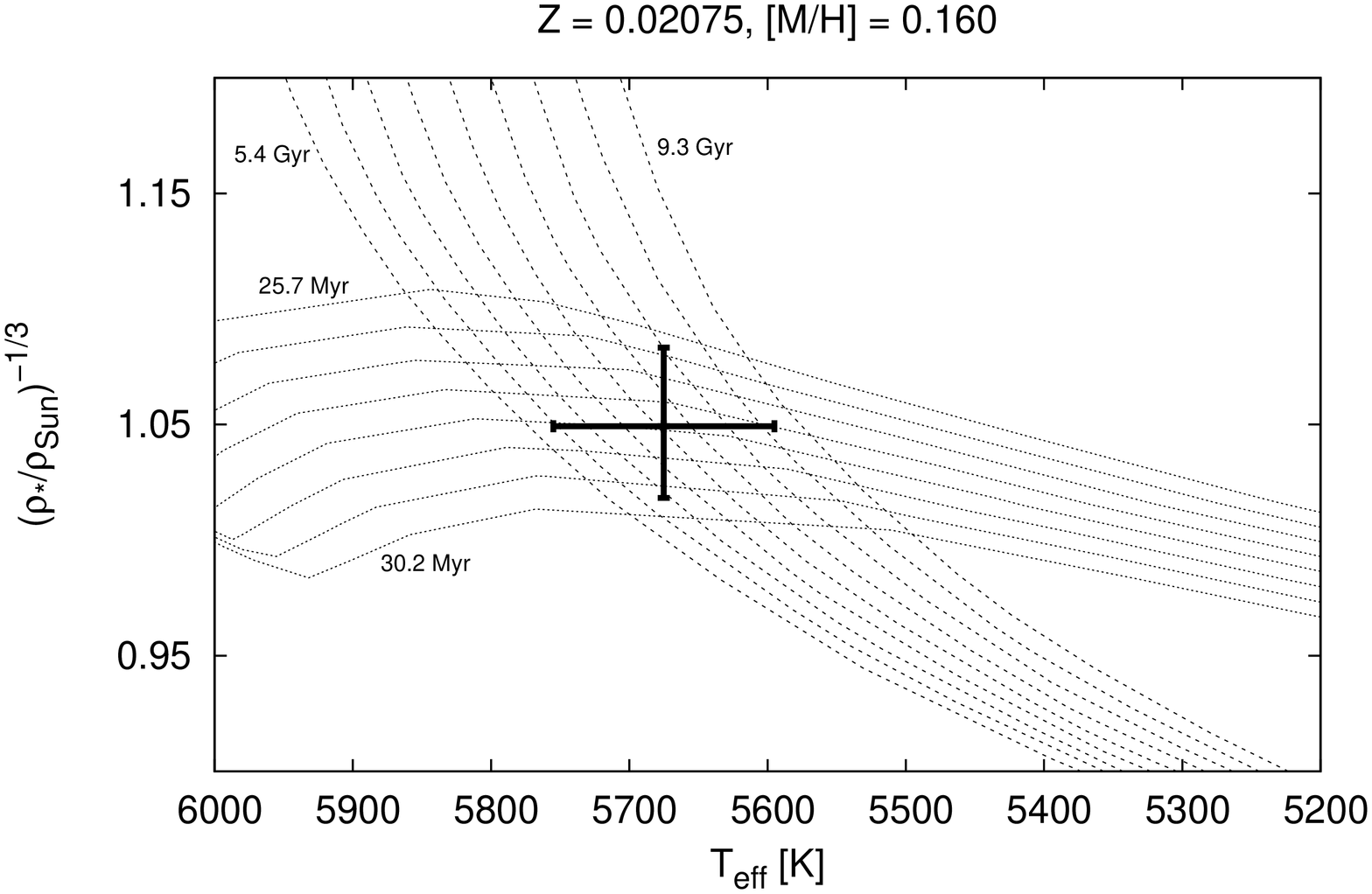}
  \caption{Position of CoRoT-12 in the $\rho_{\mathrm{A}}^{-1/3}\,-\,T_{\mathrm{eff}}$ plane. The PARSEC isochrones of metallicity [M/H]=\,0.16 for log(age)\,=\,7.41\,-\,7.48 with steps of 0.01 and log(age)\,=\,9.73\,-\,9.97 with steps of 0.03 for the young and the old age, respectively, are also shown.}
  \label{HRD_C12}
\end{minipage}
\begin{minipage}[]{0.45\textwidth}
   \includegraphics[width=0.65\textwidth,angle=270]{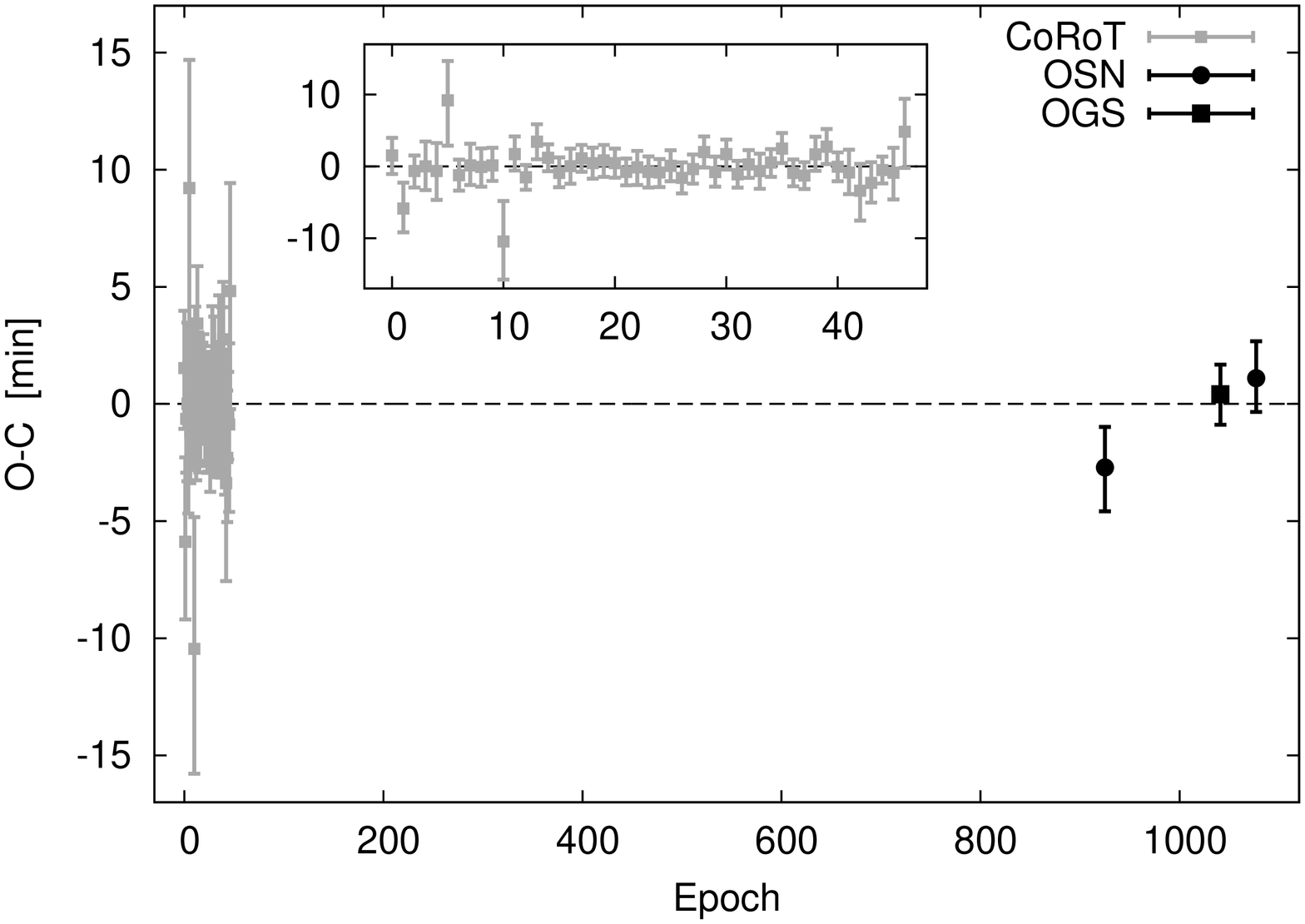}
  \caption{Same as Fig,~\ref{O_C_CoRoT5}  but for CoRoT-12\,b. The dashed line represents the updated ephemeris given in equation \ref{Elemente_CoRoT12}. }
  \label{O_C_CoRoT12}
\end{minipage}
\end{figure}

CoRoT-12\,b is a hot Jupiter that orbits its $V$=15.5\,mag, quiet and slowly rotating star in 2.83\,d. It was discovered by the \textit{CoRoT} satellite in field LRa01 which was monitored from 2007 October 24 to 2008 March 3. The transits were noticed by the `alarm mode'-pipeline after 29\,d of observations. $L_{\mathrm{3}}$ was determined through ground based photometric follow-up observations as 3.3$\pm$0.5\,\%. RV measurements were obtained with HARPS and with HIRES. CoRoT-12\,b appears to be a very low-density, inflated hot Jupiter. The slightly non-zero eccentricity was measured to be between 0.06 and 0.08. The \textit{CoRoT}-LC consist of 245\,780 unflagged measurements and contains 47 transits, 36 of them in the short cadence mode.\\ We observed three transits of CoRoT-12\,b from 2014 December to 2016 February, one at OSN and two at ESA's OGS. All three LCs together with the best-fitting model are shown in Fig.~\ref{LC_CoRoT12}. \\ The simultaneous fit of all \textit{CoRoT} and ground-based LCs (see Fig.~\ref{alltransit_phased_C12}) resulted in the system parameters given in Table~\ref{tbl:TAPmcmc1}. We calculated the stellar density and plotted CoRoT-12 in the $\rho_{\mathrm{A}}^{-1/3}\,-\,T_{\mathrm{eff}}$ plane together with the PARSEC isochrones (Fig.~\ref{HRD_C12}). As already mentioned by \citet{2010A&A...520A..97G} the age is poorly constrained. The modified HR-diagram shows overlapping isochrones of young and an old age. But since CoRoT-12 appears to be very quiet and does not show chromospheric activity or a Li absorption line, a young age seems to be unlikely. Our derived old age of log(Age)\,=\,9.79$\pm$0.26 is in agreement with the age given in \citet{2010A&A...520A..97G}. \\ As shown in Table~\ref{phys_prop_CoRoT12}, our derived physical properties of the CoRoT-12 system are in excellent agreement with the values in \citet{2010A&A...520A..97G} and \citet{2011MNRAS.417.2166S}. The transit times were used to refine the orbital ephemeris. The result is given in equation \ref{Elemente_CoRoT12} ($\chi^{2}$\,=\,26.2, reduced $\chi^{2}$\,=\,0.55):
\begin{equation}
\label{Elemente_CoRoT12}
\begin{array}{r@{.}lcr@{.}l}
T_{\mathrm{c[BJD_{TDB}]}}(E)=(2454398 & 62771 & + & E\cdot 2 & 82805268)\,\mathrm{d} \\ 
\pm 0 & 00024 &  & \pm 0 & 00000065
\end{array}
\end{equation}
The transit times and the O--C values are given in Table~\ref{CoRoT12_Transit_Times} while Fig.~\ref{O_C_CoRoT12} shows the O--C diagram calculated using the updated ephemeris. The orbital period determined by \citet{2010A&A...520A..97G} seems to be very accurate. We found a value that is less than a second higher but $\sim$20 times more precise. Although the O--C diagram of CoRoT-12 seems to show a correlated structure that was also mentioned by \citet{2010A&A...520A..97G}, the period search with \begin{scriptsize}GLS\end{scriptsize} resulted in no significant detection of TTVs. The highest peak in the periodogram with a period of $P_{\mathrm{TTV}}$\,=\,501\,$\pm$\,18\,epochs shows a FAP of 99.2\%. \citet{2010A&A...520A..97G} speculated that the structured O--C diagram may be caused by stellar rotation which could not be constrained from the \textit{CoRoT} photometry.

\begin{table}
\centering
\caption{Same as Table~\ref{phys_prop_CoRoT5} but for the CoRoT-12 system. Values derived by \citet[][G10]{2010A&A...520A..97G} and \citet[][S11]{2011MNRAS.417.2166S} are given for comparison.}
\label{phys_prop_CoRoT12}
\renewcommand{\arraystretch}{1.1}
\begin{tabular}{lr@{\,$\pm$\,}lr@{\,$\pm$\,}lr@{\,$\pm$\,}l}
\hline \hline
 Parameter & \multicolumn{2}{c}{This work} & \multicolumn{2}{c}{G10} & \multicolumn{2}{c}{S11} \\ \hline \hline
	&	 \multicolumn{6}{c}{Planetary parameters} \\ \hline 										
$R_{\mathrm{b}}$  [R$_{\mathrm{Jup}}$] 	&	1.344	&	 $^{0.074}_{0.071}$ 	&	1.44	&	0.13	&	1.350	&	0.074 \\
$M_{\mathrm{b}}$  [M$_{\mathrm{Jup}}$] 	&	0.873	&	 $^{0.081}_{0.078}$ 	&	0.917	&	 $^{0.070}_{0.065}$ 	&	0.887	&	0.077 \\
$\rho_{\mathrm{b}}$  [$\mathrm{\rho}_{\mathrm{Jup}}$] 	&	0.337	&	 $^{0.064}_{0.062}$ 	&	0.309	&	 $^{0.097}_{0.071}$ 	&	0.337	&	0.052 \\
log\,$g_{\mathrm{b}}$ 	&	3.080	&	 $^{0.047}_{0.044}$ 	&	3.043	&	 $^{0.082}_{0.080}$ 	&	3.083	&	0.047 \\
$T_{\mathrm{eq}}$ [K] 	&	1417	&	 $^{20}_{20}$ 	&	1442	&	58	&	1410	&	28 \\
$\Theta$ 	&	0.0509	&	 $^{0.0073}_{0.0071}$ 	&	 \multicolumn{2}{c}{} 			&	0.0508	&	0.0042 \\ \hline	
	&	 \multicolumn{6}{c}{Stellar parameters} \\ \hline 										
$R_{\mathrm{A}}$  [R$_{\mathrm{\odot}}$] 	&	1.049	&	 $^{0.049}_{0.047}$ 	&	1.116	&	 $^{0.096}_{0.092}$ 	&	1.046	&	0.042 \\
$M_{\mathrm{A}}$  [M$_{\mathrm{\odot}}$] 	&	1.00	&	0.10	&	1.078	&	 $^{0.077}_{0.072}$ 	&	1.018	&	0.088 \\
$\rho_{\mathrm{A}}$  [$\mathrm{\rho}_{\mathrm{\odot}}$] 	&	0.866	&	 $^{0.084}_{0.078}$ 	&	0.77	&	 $^{0.20}_{0.15}$ 	&	0.889	&	0.076 \\
log\,$g_{\mathrm{A}}$ 	&	4.396	&	 $^{0.032}_{0.030}$ 	&	4.375	&	 $^{0.065}_{0.062}$ 	&	4.407	&	0.029 \\
log$\frac{L_{\mathrm{A}}}{L_{\mathrm{\odot}}}$ 	&	0.01	&	0.07	&	 \multicolumn{2}{c}{} 			&	 \multicolumn{2}{c}{} \\		
log(Age) 	&	7.43	&	0.06	&	\multicolumn{2}{c}{} 	&	 \multicolumn{2}{c}{} \\		
	&	9.79	&	0.26	&	9.80	&	 $^{0.17}_{0.29}$		&	\multicolumn{2}{c}{}	\\ \hline 
	&	 \multicolumn{6}{c}{Geometrical parameters} \\ \hline 
	$a$  [au] 	&	 \multicolumn{2}{c}{0.0392}			&	 \multicolumn{2}{c}{0.04016}  			&	 \multicolumn{2}{c}{0.0394}  \\		
	&	 	&	0.0013	&	 	&	 $^{0.00093}_{0.00092}$	&	 	&	0.0011 \\
$i$  [$^{\circ}$] 	&	85.71	&	 $^{0.39}_{0.36}$ 	&	85.48	&	 $^{0.72}_{0.77}$ 	&	85.79	&	0.43 \\
$b$ 	&	0.604	&	 $^{0.060}_{0.057}$ 	&	0.573	&	 $^{0.027}_{0.030}$ 	&	 \multicolumn{2}{c}{} \\ \hline \hline		
\end{tabular}
\end{table}

\begin{table}
\centering
\caption{Same as Table~\ref{CoRoT5_Transit_Times} but for all transits of CoRoT-12\,b. The O--C was calculated with the ephemeris given in equation \ref{Elemente_CoRoT12}.}
\label{CoRoT12_Transit_Times}
\renewcommand{\arraystretch}{1.1}
\begin{tabular}{ccr@{\,$\pm$\,}lr@{\,$\pm$\,}l}
\hline \hline
Telescope & Epoch & \multicolumn{2}{c}{$T_{\mathrm{c}}$ [BJD$_{\mathrm{TDB}}$]} & \multicolumn{2}{c}{O--C  [min]} \\ \hline \hline
CoRoT & 0    & 2454398.6288 & $^{0.0017}_{0.0018}$  & 1.53   & $^{2.45}_{2.59}$ \\
CoRoT & 1    & 2454401.4517 & $^{0.0025}_{0.0023}$  & -5.88  & $^{3.60}_{3.31}$ \\
CoRoT & 2    & 2454404.2834 & $^{0.0015}_{0.0016}$  & -0.63  & $^{2.16}_{2.30}$ \\
CoRoT & 3    & 2454407.1119 & $^{0.0024}_{0.0023}$  & 0.01   & $^{3.46}_{3.31}$ \\
CoRoT & 4    & 2454409.9395 & $^{0.0027}_{0.0028}$  & -0.64  & $^{3.89}_{4.03}$ \\
CoRoT & 5    & 2454412.7744 & $^{0.0038}_{0.0044}$  & 9.22   & $^{5.47}_{6.34}$ \\
CoRoT & 6    & 2454415.5952 & $^{0.0015}_{0.0015}$  & -1.22  & $^{2.16}_{2.16}$ \\
CoRoT & 7    & 2454418.4242 & $^{0.0021}_{0.0019}$  & 0.14   & $^{3.02}_{2.74}$ \\
CoRoT & 8    & 2454421.2521 & $^{0.0018}_{0.0019}$  & -0.08  & $^{2.59}_{2.74}$ \\
CoRoT & 9    & 2454424.0803 & $^{0.0017}_{0.0015}$  & 0.13   & $^{2.45}_{2.16}$ \\
CoRoT & 10   & 2454426.9010 & $^{0.0039}_{0.0037}$  & -10.45 & $^{5.62}_{5.33}$ \\
CoRoT & 11   & 2454429.7375 & $^{0.0017}_{0.0016}$  & 1.71   & $^{2.45}_{2.30}$ \\
CoRoT & 12   & 2454432.5633 & $^{0.0012}_{0.0012}$  & -1.53  & $^{1.73}_{1.73}$ \\
CoRoT & 13   & 2454435.3948 & $^{0.0017}_{0.0017}$  & 3.43   & $^{2.45}_{2.45}$ \\
CoRoT & 14   & 2454438.2213 & $^{0.0013}_{0.0013}$  & 1.19   & $^{1.87}_{1.87}$ \\
CoRoT & 15   & 2454441.0479 & $^{0.0015}_{0.0014}$  & -0.90  & $^{2.16}_{2.02}$ \\
CoRoT & 16   & 2454443.8766 & $^{0.0017}_{0.0017}$  & 0.03   & $^{2.45}_{2.45}$ \\
CoRoT & 17   & 2454446.7054 & $^{0.0013}_{0.0013}$  & 1.11   & $^{1.87}_{1.87}$ \\
CoRoT & 18   & 2454449.5330 & $^{0.0015}_{0.0015}$  & 0.46   & $^{2.16}_{2.16}$ \\
CoRoT & 19   & 2454452.3613 & $^{0.0015}_{0.0016}$  & 0.82   & $^{2.16}_{2.30}$ \\
CoRoT & 20   & 2454455.1891 & $^{0.0014}_{0.0014}$  & 0.45   & $^{2.02}_{2.02}$ \\
CoRoT & 21   & 2454458.0163 & $^{0.0013}_{0.0013}$  & -0.78  & $^{1.87}_{1.87}$ \\
CoRoT & 22   & 2454460.8448 & $^{0.0016}_{0.0017}$  & -0.13  & $^{2.30}_{2.45}$ \\
CoRoT & 23   & 2454463.6724 & $^{0.0015}_{0.0015}$  & -0.78  & $^{2.16}_{2.16}$ \\
CoRoT & 24   & 2454466.5004 & $^{0.0013}_{0.0014}$  & -0.86  & $^{1.87}_{2.02}$ \\
CoRoT & 25   & 2454469.3291 & $^{0.0015}_{0.0015}$  & 0.07   & $^{2.16}_{2.16}$ \\
CoRoT & 26   & 2454472.1560 & $^{0.0015}_{0.0015}$  & -1.59  & $^{2.16}_{2.16}$ \\
CoRoT & 27   & 2454474.9849 & $^{0.0014}_{0.0014}$  & -0.37  & $^{2.02}_{2.02}$ \\
CoRoT & 28   & 2454477.8146 & $^{0.0015}_{0.0015}$  & 2.01   & $^{2.16}_{2.16}$ \\
CoRoT & 29   & 2454480.6407 & $^{0.0015}_{0.0014}$  & -0.81  & $^{2.16}_{2.02}$ \\
CoRoT & 30   & 2454483.4705 & $^{0.0014}_{0.0015}$  & 1.71   & $^{2.02}_{2.16}$ \\
CoRoT & 31   & 2454486.2966 & $^{0.0013}_{0.0013}$  & -1.10  & $^{1.87}_{1.87}$ \\
CoRoT & 32   & 2454489.1256 & $^{0.0014}_{0.0013}$  & 0.26   & $^{2.02}_{1.87}$ \\
CoRoT & 33   & 2454491.9530 & $^{0.0017}_{0.0017}$  & -0.68  & $^{2.45}_{2.45}$ \\
CoRoT & 34   & 2454494.7819 & $^{0.0013}_{0.0014}$  & 0.54   & $^{1.87}_{2.02}$ \\
CoRoT & 35   & 2454497.6113 & $^{0.0015}_{0.0014}$  & 2.48   & $^{2.16}_{2.02}$ \\
CoRoT & 36   & 2454500.4370 & $^{0.0013}_{0.0013}$  & -0.90  & $^{1.87}_{1.87}$ \\
CoRoT & 37   & 2454503.2648 & $^{0.0013}_{0.0013}$  & -1.27  & $^{1.87}_{1.87}$ \\
CoRoT & 38   & 2454506.0949 & $^{0.0017}_{0.0016}$  & 1.68   & $^{2.45}_{2.30}$ \\
CoRoT & 39   & 2454508.9237 & $^{0.0017}_{0.0017}$  & 2.76   & $^{2.45}_{2.45}$ \\
CoRoT & 40   & 2454511.7498 & $^{0.0014}_{0.0014}$  & -0.06  & $^{2.02}_{2.02}$ \\
CoRoT & 41   & 2454514.5773 & $^{0.0022}_{0.0021}$  & -0.85  & $^{3.17}_{3.02}$ \\
CoRoT & 42   & 2454517.4036 & $^{0.0026}_{0.0029}$  & -3.38  & $^{3.74}_{4.18}$ \\
CoRoT & 43   & 2454520.2324 & $^{0.0020}_{0.0019}$  & -2.30  & $^{2.88}_{2.74}$ \\
CoRoT & 44   & 2454523.0617 & $^{0.0013}_{0.0013}$  & -0.50  & $^{1.87}_{1.87}$ \\
CoRoT & 45   & 2454525.8895 & $^{0.0024}_{0.0026}$  & -0.87  & $^{3.46}_{3.74}$ \\
CoRoT & 46   & 2454528.7215 & $^{0.0032}_{0.0035}$  & 4.82   & $^{4.61}_{5.04}$ \\
OSN   & 925  & 2457014.5746 & $^{0.0012}_{0.0013}$  & -2.71  & $^{1.73}_{1.87}$ \\
OGS   & 1041 & 2457342.6308 & $^{0.0009}_{0.0009}$  & 0.41   & $^{1.27}_{1.30}$ \\
OSN   & 1077 & 2457444.4412 & $^{0.0011}_{0.0010}$  & 1.10   & $^{1.58}_{1.44}$ \\
\hline \hline
\end{tabular}
\end{table}

\section{CoRoT-18}

\begin{figure*}
  \includegraphics[width=0.475\textwidth, angle=270]{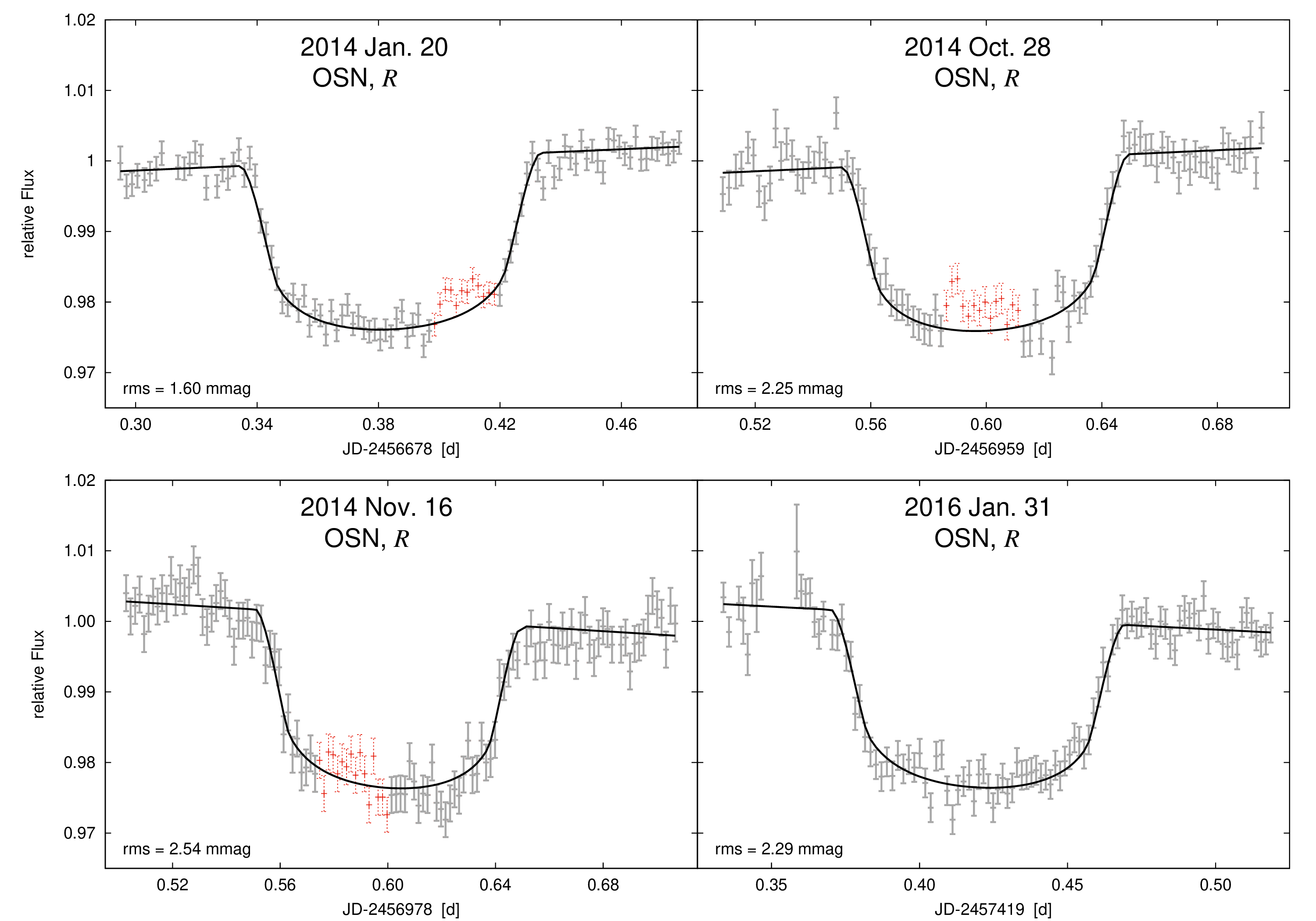}
  \caption{Same as Fig.~\ref{LC_CoRoT5} but for CoRoT-18. The parts of the LC identified as spot features by Raetz et al. (in preparation) shown here as red dashed data points were not used in the analysis (see text). The dates of observation, observatory, filter, and the $rms$ of the fit are indicated in each individual panel.}
\label{LC_CoRoT18}
\end{figure*}

\begin{figure}
\begin{minipage}[]{0.45\textwidth}
   \includegraphics[width=0.98\textwidth]{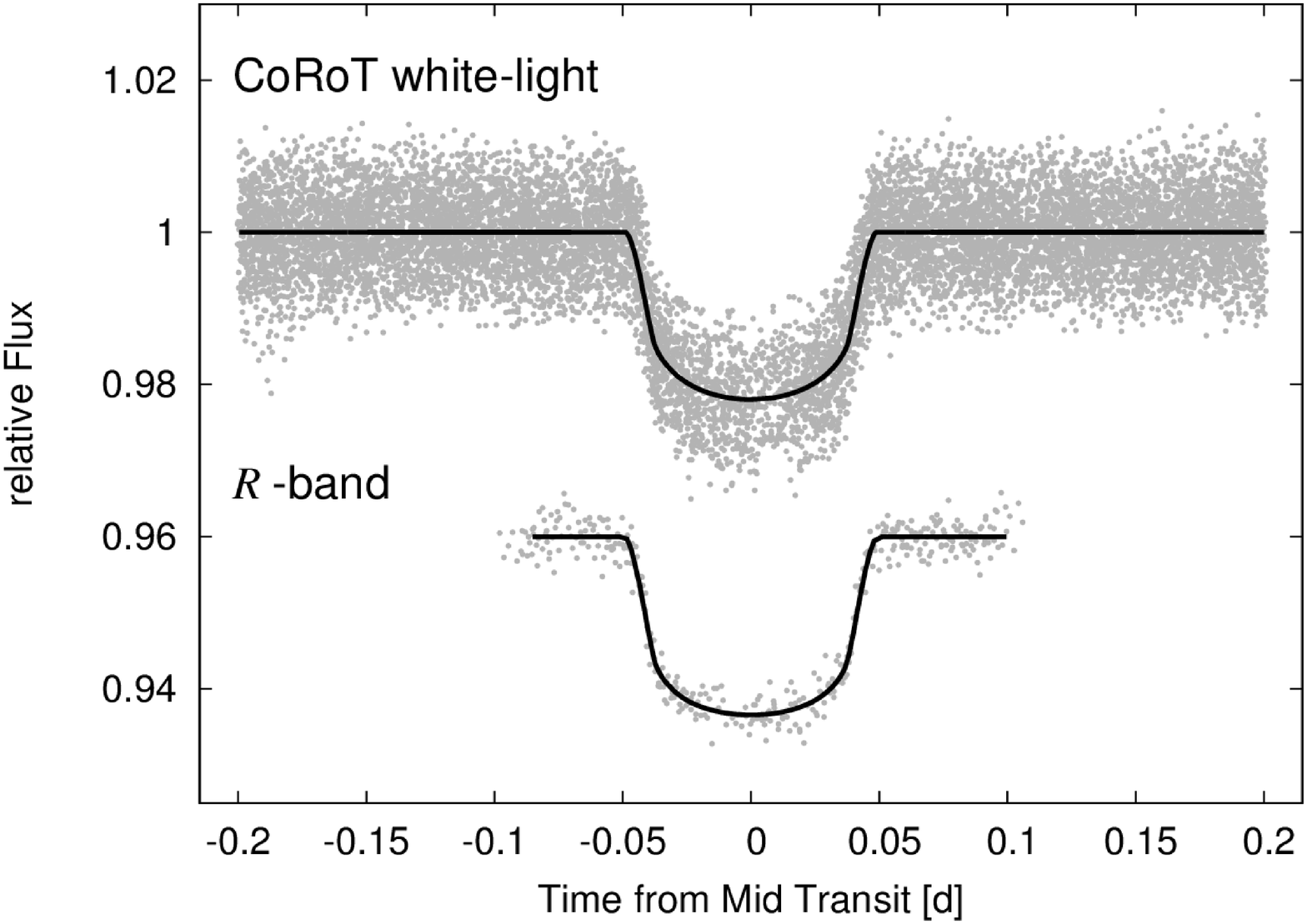}
  \caption{Phase-folded LCs of the 13 \textit{CoRoT} transits as well as of the four OSN $R$-band transits of CoRoT-18. The trend was removed before phase-folding. Overlaid are the best-fitting models obtained with TAP.}
  \label{alltransit_phased_lc_C18}
  \end{minipage}
\begin{minipage}[]{0.45\textwidth} 
   \includegraphics[width=0.7\textwidth,angle=270]{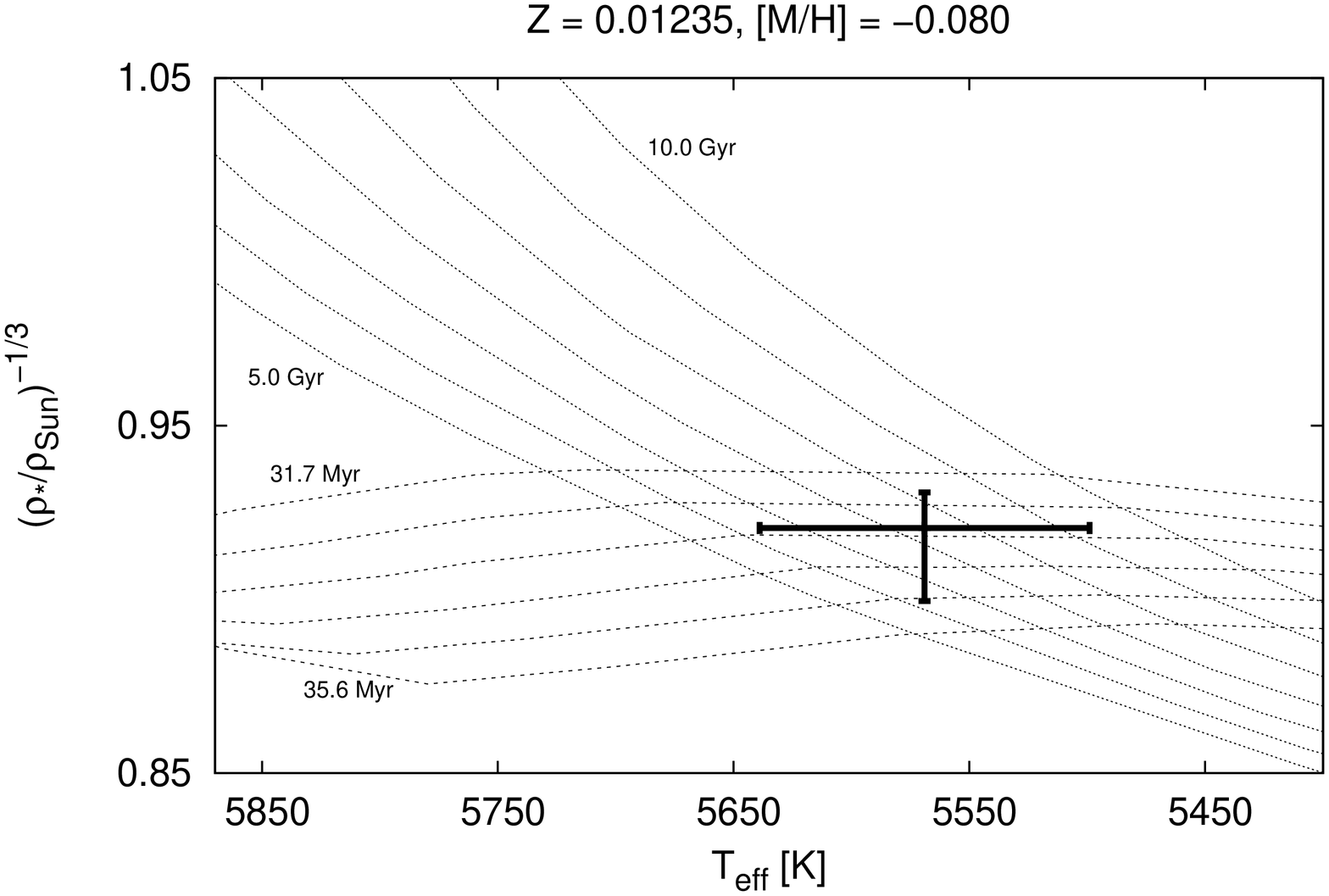}
  \caption{Position of CoRoT-18 in the $\rho_{\mathrm{A}}^{-1/3}\,-\,T_{\mathrm{eff}}$ plane. The PARSEC isochrones of metallicity [M/H]\,=\,-0.08 for log(age)\,=\,7.50\,-\,7.55 with steps of 0.01 and log(age)\,=\,9.70\,-\,10.00 with steps of 0.05 for the young and the old age, respectively, are also shown.}
  \label{HRD_C18}
\end{minipage}
\end{figure}

\begin{figure}
  \centering
  \includegraphics[width=0.28\textwidth,angle=270]{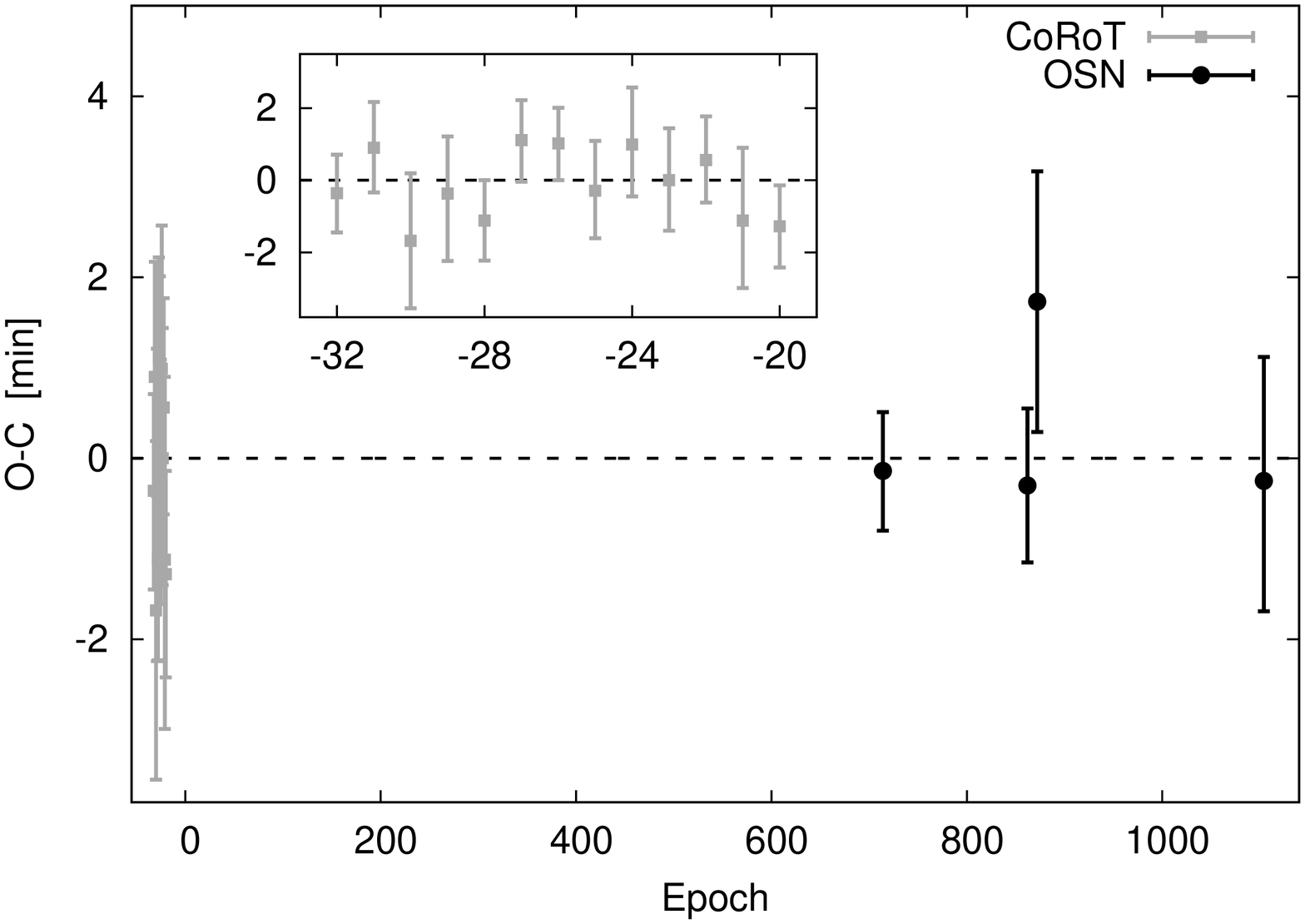}
  \caption{Same as Fig,~\ref{O_C_CoRoT5}  but for CoRoT-18\,b. The dashed line represents the updated ephemeris given in equation \ref{Elemente_CoRoT18}. }
  \label{O_C_CoRoT18}
\end{figure}

CoRoT-18\,b was detected in the field SRa03 that was observed by \textit{CoRoT} from 2010 March 5 to 29 \citep{2011A&A...533A.130H}. It is a massive hot Jupiter that orbits its G9V host star in $\sim$1.9\,d. Its eccentricity is slightly non-zero (e\,$<$\,0.08) and therefore the planet also belongs to the group of massive planets on elliptical orbits. \citet{2013A&A...550A..67P} reported a statistically marginal detection of a secondary eclipse near a phase of 0.47 which corresponds to e\,=\, 0.10\,$\pm$\,0.04, and, hence, confirms the non-zero eccentricity. The ground-based LC of CoRoT-18 presented in \citet{2011A&A...533A.130H} revealed a brightness bump in-transit that could arise from a starspot crossing, therefore supporting the hypothesis of CoRoT-18 being a young star. However, the analysis of CoRoT-18 yielded inconsistent age determinations. While the stellar activity, lithium abundance, and stellar spin point to a young age, the evolutionary tracks do not exclude very old ages.\\ Based on lucky imaging observations in two different filters, \citet{2016A&A...589A..58E} suggested the existence of a possible companion candidate to CoRoT-18 at a separation of at least 8000\,au. No definitive conclusion could be drawn due to large measurement uncertainties. \\ The cadence of the \textit{CoRoT} measurements was 32\,s throughout the observations. After removing all flagged entries we were left with 56\,823 data points. The LC includes in total 13 transit events. The contamination factor was found to be $L_{\mathrm{3}}\,=\,2.0\pm0.1\%$ by \citet{2011A&A...533A.130H}.\\ We observed four transit events in 2014 and 2016 at the OSN. The ground-based as well as the \textit{CoRoT} LCs of CoRoT-18 show brightness bumps that could be attributed to stellar spots. Stellar activity complicates transit modelling due to the non-homogeneous brightness distribution on the stellar surface \citep[e.g.][]{2009A&A...505.1277C,2013A&A...556A..19O}. If occulted and unocculted spots outside the transit path are not correctly modelled, systematic errors in the determination of the system parameters will arise. The detailed spot modelling for CoRoT-18 is discussed in Raetz et al. (in preparation). Before the simultaneous transit fitting of all \textit{CoRoT} white light and the ground-based \textit{R}-band transits with \begin{scriptsize}TAP\end{scriptsize} we removed all parts of the LCs where spot-features were identified by Raetz et al. (in preparation). The ground-based LCs with the best-fitting model, the simultaneous fit of all \textit{CoRoT} and ground-based LCs and the resulting system parameters are given in Figs.~\ref{LC_CoRoT18} and \ref{alltransit_phased_lc_C18} and Table~\ref{tbl:TAPmcmc1}, respectively.\\ By plotting CoRoT-18 in the $\rho_{\mathrm{A}}^{-1/3}\,-\,T_{\mathrm{eff}}$ plane (Fig.~\ref{HRD_C18}), we confirm the finding of \citet{2011A&A...533A.130H} that CoRoT-18 is consistent with very young ($\sim$\,33\,Myr) and old ($\sim$\,7\,Gyr) ages. The derived physical properties that are summarized in Table~\ref{phys_prop_CoRoT18} agree, on average within $\sim$1.1\,$\sigma$, with the values of \citet{2011A&A...533A.130H} and \citet{2012MNRAS.426.1291S}. We found the largest deviations from the \citet{2011A&A...533A.130H} values for the inclination ($\sim$2.4\,$\sigma$), the impact parameter($\sim$2.8\,$\sigma$), and the stellar density ($\sim$1.9\,$\sigma$). These discrepancies most likely arise from the different treatment of the stellar activity.\\ We used the transit times derived by the simultaneous transit modelling with \begin{scriptsize}TAP\end{scriptsize} of the spot removed LCs to refine the ephemeris. Our OSN observations were carried out 4-6 yr after the \textit{CoRoT} discovery. Using the original ephemeris of \citet{2011A&A...533A.130H} the calculated transit times deviate from the observed ones by up to $\sim$34\,min. Within total 17 mid-transit times, we have been able to refine the orbital elements and improve their precision. The result is given in equation \ref{Elemente_CoRoT18} ($\chi^{2}$\,=\,8.4, reduced $\chi^{2}$\,=\,0.56): 
\begin{equation}
\label{Elemente_CoRoT18}
\begin{array}{r@{.}lcr@{.}l}
T_{\mathrm{c[BJD_{TDB}]}}(E)=(2455321 & 72565 & + & E\cdot 1 & 9000900)\,\mathrm{d} \\ 
\pm 0 & 00024 &  & \pm 0 & 0000005
\end{array}
\end{equation}
The orbital period $P$ is 1.8\,s longer and six times more precise than the one given in \citet{2011A&A...533A.130H}. The transit times and O--C values are given in  Table~\ref{CoRoT18_Transit_Times} while Fig.~\ref{O_C_CoRoT18} shows the resulting O--C diagram. We could not find indications for TTVs. \begin{scriptsize}GLS\end{scriptsize} resulted in a period of $P_{\mathrm{TTV}}$\,=\,75.0\,$\pm$\,0.2\,epochs with an FAP of 99.8\%.

\begin{table}
\centering
\caption{Same as Table~\ref{phys_prop_CoRoT5} but for the CoRoT-18 system. Values derived by \citet[][H11]{2011A&A...533A.130H} and \citet[][S12]{2012MNRAS.426.1291S} are given for comparison.}
\label{phys_prop_CoRoT18}
\renewcommand{\arraystretch}{1.1}
\begin{tabular}{lr@{\,$\pm$\,}lr@{\,$\pm$\,}lr@{\,$\pm$\,}l}
\hline \hline
 Parameter & \multicolumn{2}{c}{This work} & \multicolumn{2}{c}{H11} & \multicolumn{2}{c}{S12} \\ \hline \hline
& \multicolumn{6}{c}{Planetary parameters} \\ \hline 
$R_{\mathrm{b}}$  [R$_{\mathrm{Jup}}$] & 1.146 & $^{0.039}_{0.048}$ & 1.31 & 0.18 & 1.251 & 0.083 \\
$M_{\mathrm{b}}$  [M$_{\mathrm{Jup}}$] & 3.30 & $^{0.19}_{0.19}$ & 3.47 & 0.38 & 3.27 & 0.17 \\
$\rho_{\mathrm{b}}$  [$\mathrm{\rho}_{\mathrm{Jup}}$] & 2.06 & $^{0.24}_{0.29}$ & 1.65 & 0.60 & 1.56 & 0.30 \\
log\,$g_{\mathrm{b}}$ & 3.797 & $^{0.021}_{0.030}$ & \multicolumn{2}{c}{}  & 3.714 & 0.055 \\
$T_{\mathrm{eq}}$ [K] & 1487 & $^{19}_{19}$ & 1550 & 90 & 1490 & 45 \\ 
$\Theta$ & 0.189 & $^{0.019}_{0.020}$ & \multicolumn{2}{c}{} & 0.173 & 0.012 \\ \hline
& \multicolumn{6}{c}{Stellar parameters} \\ \hline 
$R_{\mathrm{A}}$  [R$_{\mathrm{\odot}}$] & 0.883 & $^{0.025}_{0.031}$ & 1.00 & 0.13 & 0.924 & 0.057 \\
$M_{\mathrm{A}}$  [M$_{\mathrm{\odot}}$] & 0.88 & 0.07 & 0.95 & 0.15 & 0.861 & 0.059 \\
$\rho_{\mathrm{A}}$  [$\mathrm{\rho}_{\mathrm{\odot}}$] & 1.28 & $^{0.04}_{0.09}$ & 0.96 & 0.17 & 1.09 & 0.16 \\ 
log\,$g_{\mathrm{A}}$ & 4.491 & $^{0.015}_{0.023}$ & 4.4 & 0.1 & 4.442 &  0.043 \\
log$\frac{L_{\mathrm{A}}}{L_{\mathrm{\odot}}}$ & -0.17 & 0.06 & \multicolumn{2}{c}{} & \multicolumn{2}{c}{} \\
log(Age) & 7.50 & 0.04 & \multicolumn{2}{c}{} & \multicolumn{2}{c}{} \\
  & 9.84 & 0.26 & \multicolumn{2}{c}{} & \multicolumn{2}{c}{} \\\hline 
& \multicolumn{6}{c}{Geometrical parameters} \\ \hline 
$a$  [au] & \multicolumn{2}{c}{0.0288}  & \multicolumn{2}{c}{0.0295}  & \multicolumn{2}{c}{0.0286}  \\
& & 0.0008 & & 0.0016 & & 0.0007 \\
$i$  [$^{\circ}$] & 89.9 & $^{1.6}_{1.6}$ & 86.5 & $^{1.4}_{0.9}$ & 86.8 & 1.7 \\
$b$ & 0.01 & $^{0.20}_{0.20}$ & 0.40 & $^{0.08}_{0.14}$ & \multicolumn{2}{c}{} \\ \hline \hline
\end{tabular}
\end{table}

\begin{table}
\centering
\caption{Same as Table~\ref{CoRoT5_Transit_Times} but for all transits of CoRoT-18\,b. The O--C was calculated with the ephemeris given in equation \ref{Elemente_CoRoT18}. $T_{\mathrm{c}}$: mid-transit time of the spot removed LCs.}
\label{CoRoT18_Transit_Times}
\renewcommand{\arraystretch}{1.1}
\begin{tabular}{ccr@{\,$\pm$\,}lr@{\,$\pm$\,}l}
\hline \hline
 Telescope & Epoch & \multicolumn{2}{c}{$T_{\mathrm{c}}$ [BJD$_{\mathrm{TDB}}$]} & \multicolumn{2}{c}{O--C  [min]} \\ \hline \hline
CoRoT & -32  &  2455260.922523 &  $^{0.00074}_{0.00076}$ & -0.3	&  $^{1.07}_{1.09}$ \\
CoRoT & -31  &  2455262.823483 &  $^{0.00088}_{0.00086}$ &  0.9	&  $^{1.27}_{1.24}$ \\
CoRoT & -30  &  2455264.721783 &  $^{0.00130}_{0.00130}$ & -1.6	&  $^{1.87}_{1.87}$ \\
CoRoT & -29  &  2455266.622783 &  $^{0.00110}_{0.00130}$ & -0.3	&  $^{1.58}_{1.87}$ \\
CoRoT & -28  &  2455268.522353 &  $^{0.00078}_{0.00077}$ & -1.1	&  $^{1.12}_{1.11}$ \\
CoRoT & -27  &  2455270.423993 &  $^{0.00077}_{0.00080}$ &  1.1	&  $^{1.11}_{1.15}$ \\
CoRoT & -26  &  2455272.324023 &  $^{0.00069}_{0.00071}$ &  1.0	&  $^{0.99}_{1.02}$ \\
CoRoT & -25  &  2455274.223202 &  $^{0.00096}_{0.00092}$ & -0.2	&  $^{1.38}_{1.32}$ \\
CoRoT & -24  &  2455276.124182 &  $^{0.00110}_{0.00100}$ &  0.9	&  $^{1.58}_{1.44}$ \\
CoRoT & -23  &  2455278.023582 &  $^{0.00100}_{0.00097}$ &  0.0	&  $^{1.44}_{1.40}$ \\
CoRoT & -22  &  2455279.924062 &  $^{0.00084}_{0.00082}$ &  0.5	&  $^{1.21}_{1.18}$ \\
CoRoT & -21  &  2455281.822982 &  $^{0.00140}_{0.00130}$ & -1.1	&  $^{2.02}_{1.87}$ \\
CoRoT & -20  &  2455283.722962 &  $^{0.00079}_{0.00079}$ & -1.2	&  $^{1.14}_{1.14}$ \\
OSN   & 714  &  2456678.389800 &  $^{0.00045}_{0.00046}$ & -0.1	&  $^{0.65}_{0.66}$ \\
OSN   & 862  &  2456959.603001 &  $^{0.00059}_{0.00059}$ & -0.3	&  $^{0.85}_{0.85}$ \\
OSN   & 872  &  2456978.605317 &  $^{0.00100}_{0.00100}$ &  1.7	&  $^{1.44}_{1.44}$ \\ 
OSN   & 1104 &  2457419.424813 &  $^{0.00095}_{0.00100}$ & -0.2	&  $^{1.37}_{1.44}$ \\
\hline \hline
\end{tabular}
\end{table}

\section{CoRoT-20}

\begin{figure*}
  \includegraphics[width=0.235\textwidth, angle=270]{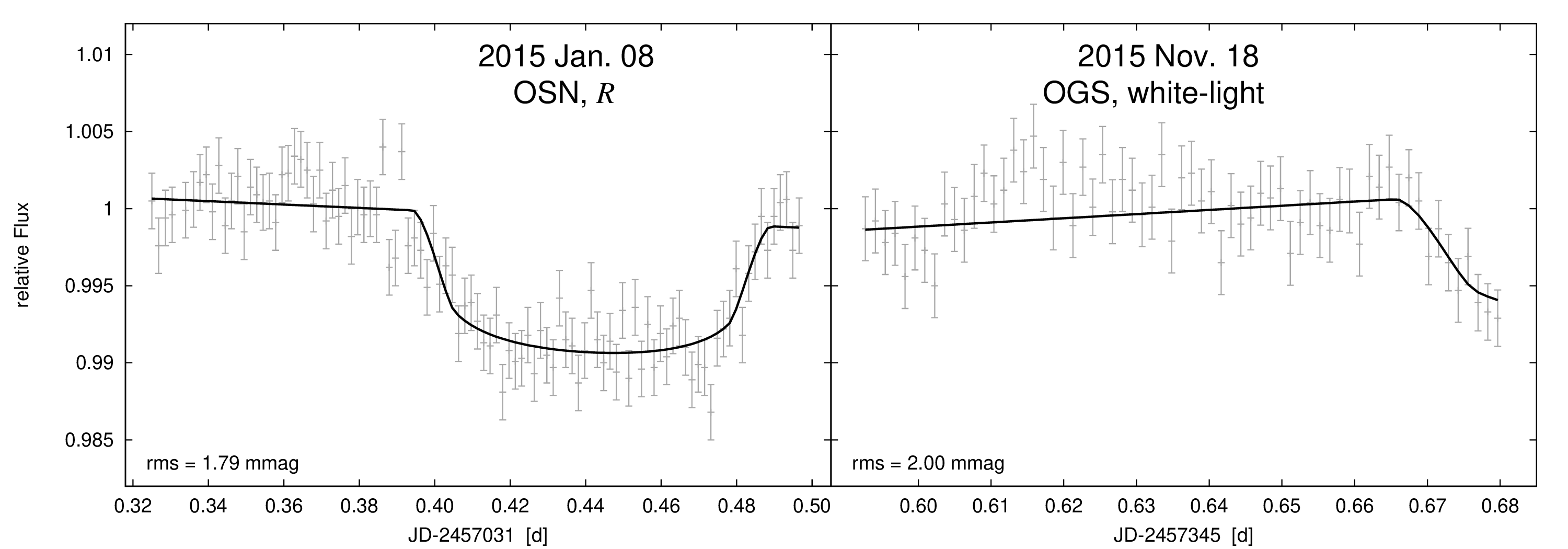}
  \caption{Same as Fig.~\ref{LC_CoRoT5} but for CoRoT-20. The dates of observation, observatory, filter, and the $rms$ of the fit are indicated in each individual panel.}
\label{LC_CoRoT20}
\end{figure*}

\begin{figure}
  \includegraphics[width=0.45\textwidth]{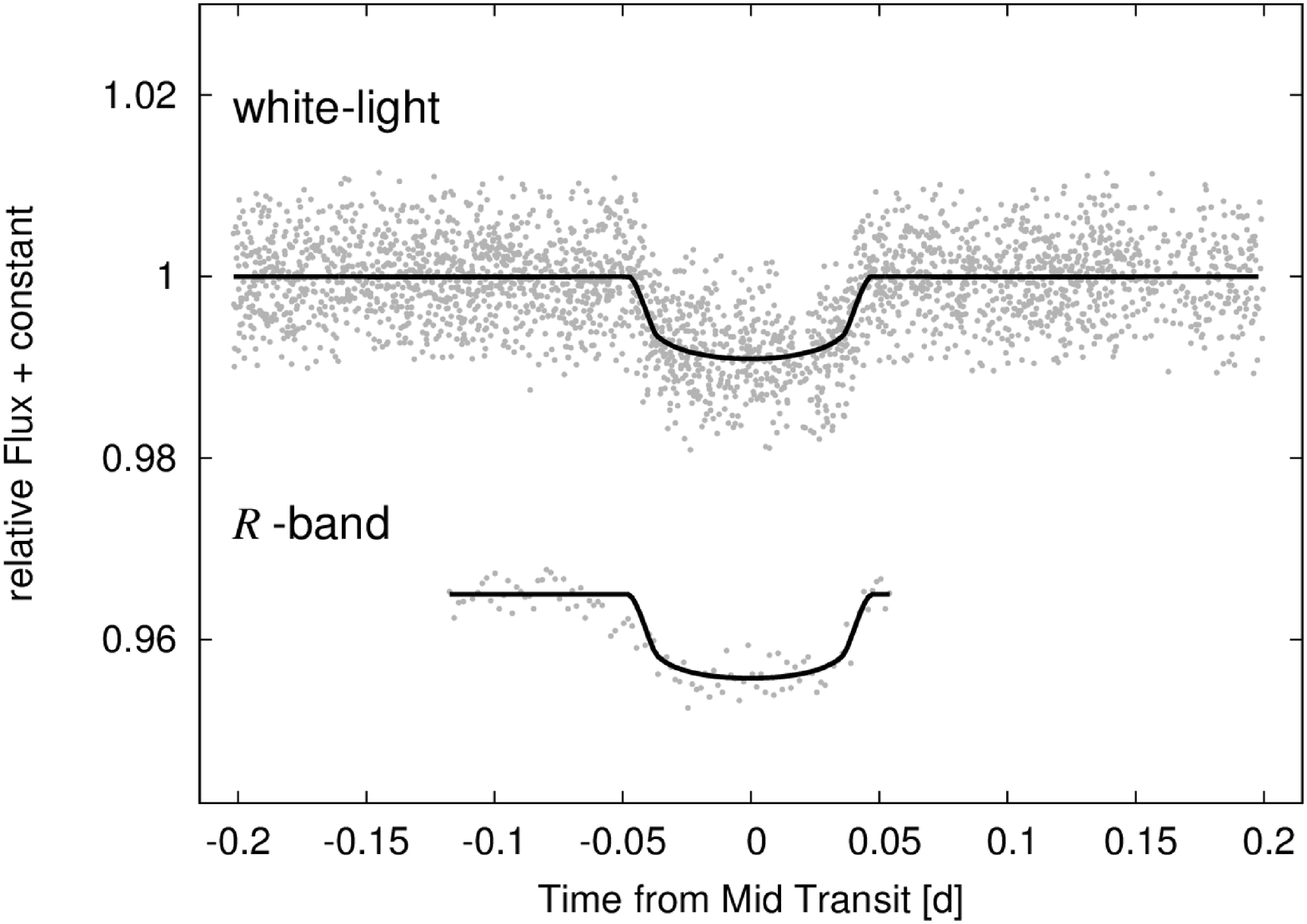}
  \caption{Phase-folded LCs of all three \textit{CoRoT} transits as well as our own transits of CoRoT-20. The trend was removed before phase-folding. Overlaid are the best-fitting models obtained with TAP. }
  \label{alltransit_phased_C20}
\end{figure}

\begin{figure}
\begin{minipage}[]{0.45\textwidth}
  \includegraphics[width=0.65\textwidth,angle=270]{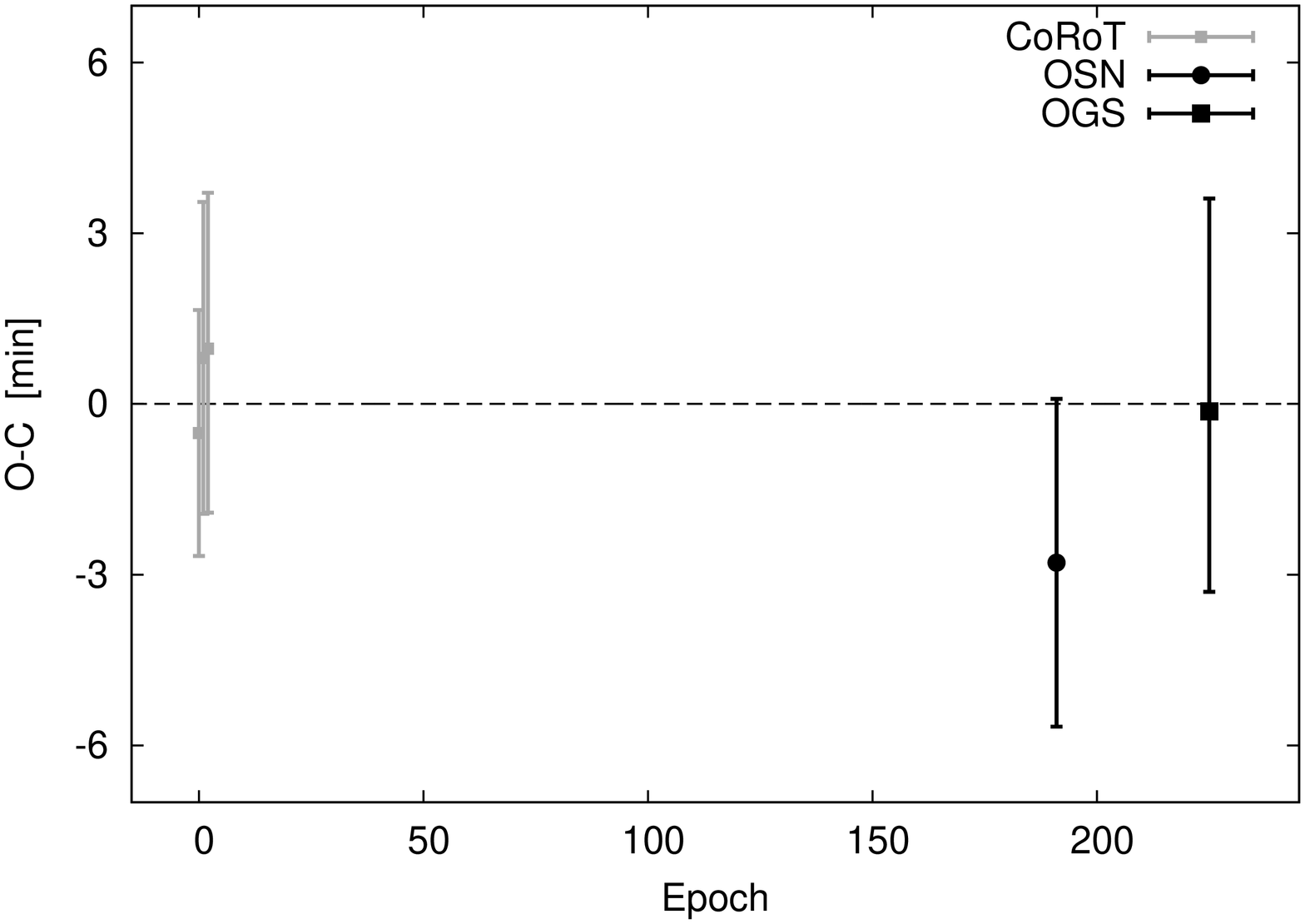}
  \caption{Same as Fig.~\ref{O_C_CoRoT5}  but for CoRoT-20\,b. The dashed line represents the updated ephemeris given in equation \ref{Elemente_CoRoT20}. }
  \label{O_C_CoRoT20}
\end{minipage}
\begin{minipage}[]{0.45\textwidth}
  \includegraphics[width=0.65\textwidth,angle=270]{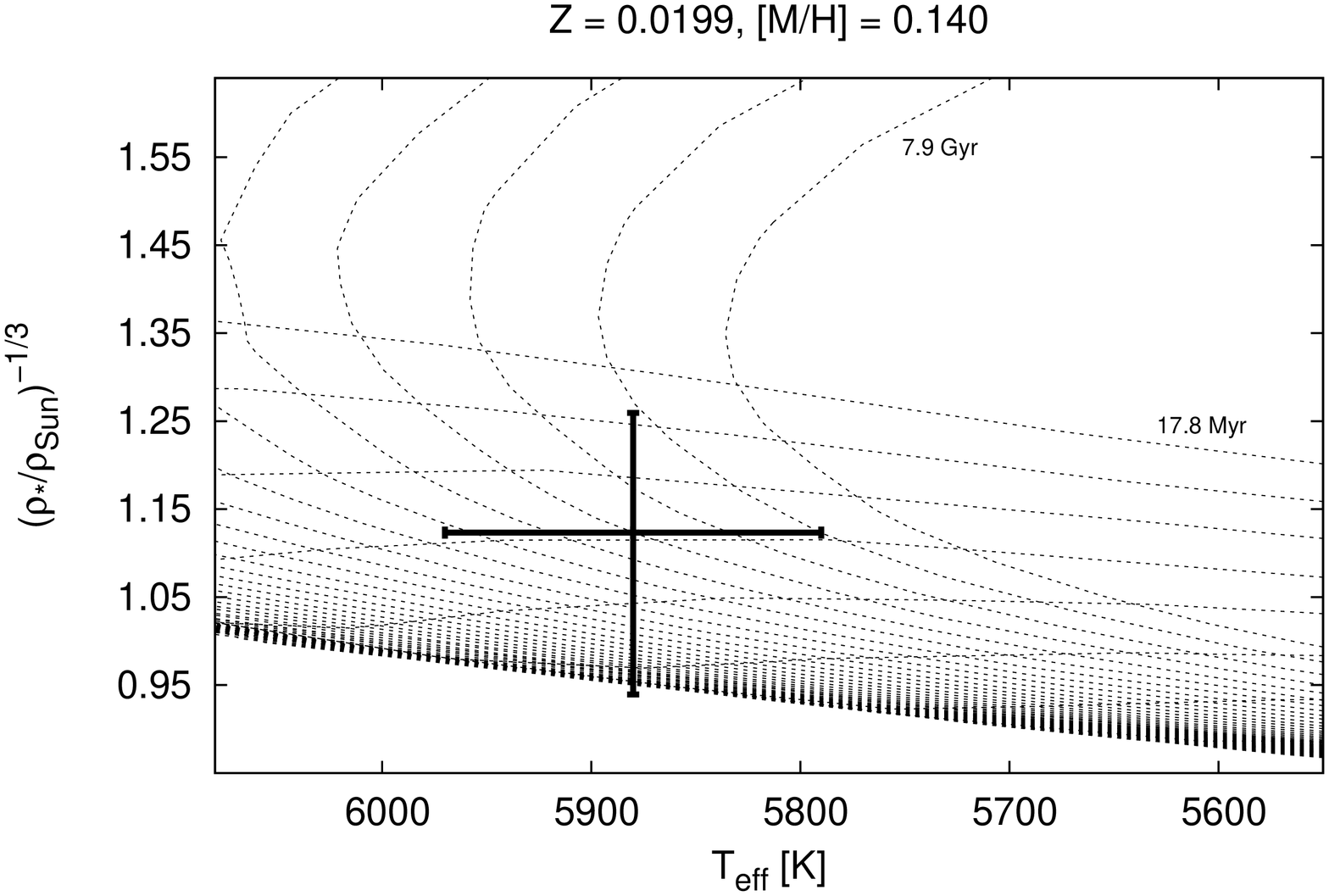}
  \caption{Position of CoRoT-20 in the $\rho_{\mathrm{A}}^{-1/3}\,-\,T_{\mathrm{eff}}$ plane. The PARSEC isochrones of metallicity [M/H]=\,0.14 for log(age)\,=\,7.25\,-\,9.90 with steps of 0.05 are also shown.}
  \label{HRD_C20}
\end{minipage}
\end{figure}

CoRoT-20\,b is another hot Jupiter that was discovered in the \textit{CoRoT} field SRa03 which was monitored for $\sim$24.3\,d starting on 2010 March 1 \citep{2012A&A...538A.145D}. The $\sim$10\,mmag  deep transit event was detected by the `alarm mode'-pipeline which triggered ground-based follow-up observations. Photometric transit observations and RV measurements were carried out at the WISE observatory and with HARPS, SOPHIE, and FIES at the NOT, respectively. The planet orbits its G2-type dwarf with an orbital period of 9.24\,d and an eccentricity of 0.56. CoRoT-20\,b belongs to the most compact planets known so far. It is an unusual and, hence, a very interesting object as it populates the border of the gap between hot Jupiters and very massive hot Jupiters in the period-mass diagram for close-in exoplanets \citep[P\,$<$\,10\,d and M\,$<$\,15 M$_{\mathrm{Jup}}$, see Fig.~8 in][]{2015MNRAS.451.4139R}. Because of its relatively long period only three transit events could be observed during the SR. Images of the area around the star showed that CoRoT-20 is rather isolated resulting in a very low $L_{\mathrm{3}}$ of less than 0.6\%. The failure of the CoRoT DPU No.1, in 2009 March, reduced the total number of stars observed, while allowing to study more of them with the higher sampling rate. Therefore, all data of CoRoT-20 were acquired in short cadence mode. The white-light LC including the three transit events consists of 56\,860 unflagged data points.\\ We observed two transits of CoRoT-20\,b in 2015 January and November, one at the OSN and one partial event at ESA's OGS (see Fig.~\ref{LC_CoRoT20}). While the slight eccentricity of our other targets only marginal affected the transit shape, the eccentricity of 0.56 for CoRoT-20\,b cannot be neglected in the simultaneous transit modelling. We fixed the eccentricity to the value given in \citet{2012A&A...538A.145D}. The result of the joint modelling of space- and ground-based LCs is given in Table~\ref{tbl:TAPmcmc1} and shown in Fig.~\ref{alltransit_phased_C20}. The transit times obtained from the transit fitting given in Table~\ref{CoRoT20_Transit_Times} allowed us to re-determine the ephemeris. The result is given in equation \ref{Elemente_CoRoT20} ($\chi^{2}$\,=\,0.69, reduced $\chi^{2}$\,=\,0.23):
\begin{equation}
\label{Elemente_CoRoT20}
\begin{array}{r@{.}lcr@{.}l}
T_{\mathrm{c[BJD_{TDB}]}}(E)=(2455266 & 0016 & + & E\cdot 9 & 243180)\,\mathrm{d} \\ 
\pm 0 & 0010 &  & \pm 0 & 000009
\end{array}
\end{equation}
Fig.~\ref{O_C_CoRoT20} shows the O--C diagram created with the updated ephemeris. The orbital period is $\sim$28\,s higher and 33 times more precise than the one given in \citet{2012A&A...538A.145D}. With only five measurements we could not find any TTVs.\\ The physical properties are summarized in Table~\ref{phys_prop_CoRoT20}. The parameters are in good agreement with the ones of \citet{2012A&A...538A.145D}. The largest deviation we found is 1.3$\sigma$ for the planetary density.\\ CoRoT-20 appears to be a quite star as its LC does not show any features. In addition, the spectra show no signs of chromospheric activity. Because of the measurable Li-line CoRoT-20 is likely a young star in the last stages of the pre-main-sequence phase \citep{2012A&A...538A.145D}. Our measurements confirm the age estimate of \citet{2012A&A...538A.145D} but are less precise. The modified HR-diagram together with the PARSEC isochrones can be found in Fig.~\ref{HRD_C20}.

\begin{table}
\centering
\caption{Same as Table~\ref{CoRoT5_Transit_Times} but for all transits of CoRoT-20\,b. The O--C was calculated with the ephemeris given in equation \ref{Elemente_CoRoT20}.}
\label{CoRoT20_Transit_Times}
\renewcommand{\arraystretch}{1.1}
\begin{tabular}{ccr@{\,$\pm$\,}lr@{\,$\pm$\,}l}
\hline \hline
Telescope & Epoch & \multicolumn{2}{c}{$T_{\mathrm{c}}$ [BJD$_{\mathrm{TDB}}$]} & \multicolumn{2}{c}{O--C  [min]} \\ \hline \hline
CoRoT & 0    & 2455266.0011 & $^{0.0015}_{0.0015}$  & -0.51 & $^{2.16}_{2.16}$ \\
CoRoT & 1    & 2455275.2452 & $^{0.0019}_{0.0019}$  & 0.81  & $^{2.74}_{2.74}$ \\
CoRoT & 2    & 2455284.4885 & $^{0.0019}_{0.0020}$  & 0.97  & $^{2.74}_{2.88}$ \\
OSN & 191    & 2457031.4480 & $^{0.0020}_{0.0020}$  & -2.79 & $^{2.88}_{2.88}$ \\
OGS & 225    & 2457345.7182 & $^{0.0026}_{0.0022}$  & -0.13 & $^{3.74}_{3.17}$ \\
\hline \hline
\end{tabular}
\end{table}

\begin{table}
\centering
\caption{Same as Table~\ref{phys_prop_CoRoT5} but for the CoRoT-20 system. Values derived by \citet[][D12]{2012A&A...538A.145D} and \citet[][S12]{2012MNRAS.426.1291S} are given for comparison.}
\label{phys_prop_CoRoT20}
\renewcommand{\arraystretch}{1.1}
\begin{tabular}{lr@{\,$\pm$\,}lr@{\,$\pm$\,}lr@{\,$\pm$\,}l}
\hline \hline
 Parameter & \multicolumn{2}{c}{This work} & \multicolumn{2}{c}{D12} & \multicolumn{2}{c}{S12} \\ \hline \hline
	&	 \multicolumn{6}{c}{Planetary parameters} \\ \hline 										
$R_{\mathrm{b}}$  [R$_{\mathrm{Jup}}$] 	&	1.00	&	 $^{0.18}_{0.21}$ 	&	0.84	&	0.04	&	1.16	&	0.26 \\
$M_{\mathrm{b}}$  [M$_{\mathrm{Jup}}$] 	&	4.14	&	 $^{0.36}_{0.3}$ 	&	4.24	&	0.23	&	5.06	&	0.36 \\
$\rho_{\mathrm{b}}$  [$\mathrm{\rho}_{\mathrm{Jup}}$] 	&	3.9	&	 $^{2.1}_{2.4}$	&	6.67	&	0.83	&	3.0	&	2.5 \\
log\,$g_{\mathrm{b}}$ 	&	4.01	&	 $^{0.15}_{0.18}$ 	&	 \multicolumn{2}{c}{} 			&	3.968	&	0.215 \\
$T_{\mathrm{eq}}$ [K] 	&	1024	&	 $^{16}_{16}$ 	&	1002	&	24	&	1100	&	150 \\
$\Theta$ 	&	0.67	&	 $^{0.16}_{0.17}$ 	&	 \multicolumn{2}{c}{} 			&	0.70	&	0.17 \\ \hline 
	&	 \multicolumn{6}{c}{Stellar parameters} \\ \hline 										
$R_{\mathrm{A}}$  [R$_{\mathrm{\odot}}$] 	&	1.16	&	 $^{0.15}_{0.20}$ 	&	1.02	&	0.05	&	1.34	&	0.37 \\
$M_{\mathrm{A}}$  [M$_{\mathrm{\odot}}$] 	&	1.10	&	0.1	&	1.14	&	0.08	&	1.11	&	0.01 \\
$\rho_{\mathrm{A}}$  [$\mathrm{\rho}_{\mathrm{\odot}}$] 	&	0.71	&	 $^{0.26}_{0.35}$	&	1.071	&	 $^{0.032}_{0.037}$ 	&	0.46	&	0.48 \\
log\,$g_{\mathrm{A}}$ 	&	4.35	&	 $^{0.11}_{0.14}$ 	&	4.20	&	0.15	&	4.23	&	0.24 \\
log$\frac{L_{\mathrm{A}}}{L_{\mathrm{\odot}}}$ 	&	0.17	&	0.18	&	 \multicolumn{2}{c}{} 			&	 \multicolumn{2}{c}{} \\		
log(Age) 	&	8.6	&	1.4	&	8.00	&	 $^{0.95}_{0.22}$ 	&	 \multicolumn{2}{c}{} \\ \hline 		
	&	 \multicolumn{6}{c}{Geometrical parameters} \\ \hline 										
$a$  [au] 	&	 \multicolumn{2}{c}{0.0891}			&	 \multicolumn{2}{c}{0.0902}  			&	 \multicolumn{2}{c}{0.0892}  \\		
	&	 	&	0.0038	&	 	&	0.0021	&	 	&	0.0028 \\
$i$  [$^{\circ}$] 	&	85.9	&	 $^{2.5}_{2.2}$ 	&	88.21	&	0.53	&	83.5	&	3.8 \\
$b$ 	&	0.6	&	 $^{0.4}_{0.3}$ 	&	0.26	&	0.08	&	 \multicolumn{2}{c}{} \\ \hline \hline		
\end{tabular}
\end{table}


\section{CoRoT-27}

\begin{figure*}
  \includegraphics[width=0.235\textwidth, angle=270]{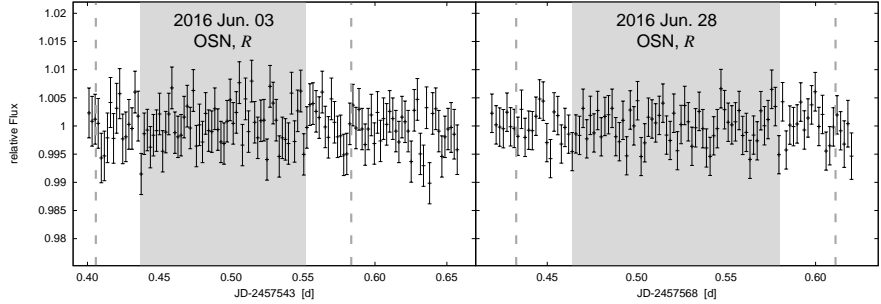}
  \caption{LCs of CoRoT-27. The grey area indicates the predicted transit window. The dashed lines give the transit beginning and end including the uncertainties of the ephemeris given in \citet{2014A&A...562A.140P}. The transit event was not detected. The date of observation, observatory, and filter are indicated in each individual panel.}
\label{LC_CoRoT27}
\end{figure*}

CoRoT-27\,b is a very massive ($M=10.39\pm0.55\,M_{\mathrm{Jup}}$) transiting planet on a 3.58\,d orbit around a 4.2 Gyr-old G2 star \citep{2014A&A...562A.140P}. It was detected in the field LRc08 that was observed continuously by \textit{CoRoT} for 83.5\,d (from 2011 July 8 to 2011 September 30). It belongs, like CoRoT-20\,b, to the densest exoplanets known so far. Although  many of the so-called hot super-Jupiters have elliptical orbits, the 13 RV measurements of CoRoT-27\,b obtained with HARPS in summer 2012 by \citet{2014A&A...562A.140P} do not indicate a significant non-zero eccentricity. Furthermore, massive close-in planets are mostly found around F-type stars and only rarely around G-stars, as it is the case for CoRoT-27\,b. This makes CoRoT-27\,b an important target to constrain formation, migration, and evolution of gas giant planets.\\ We scheduled two transit observations of CoRoT-27\,b in 2016 June with OSN. In both cases we obtained good quality LCs covering the whole predicted transit window \citep[predicted using the ephemeris of][]{2014A&A...562A.140P} including out-of-transit data before and after the assumed transit time. In none of the LCs we could detect the transit event. As shown in Fig.~\ref{LC_CoRoT27} the $\sim$1\% deep transit event should have easily been detected. The dashed lines in Fig.~\ref{LC_CoRoT27} give the range of the transit beginning and end times expected from the uncertainties in the ephemeris of \citet{2014A&A...562A.140P}. The non-detection indicates that the original determined orbital period was not accurate enough to predict the transit event 5\,yr later. Our LCs provide a lower limit for the deviation from the predicted transit time. The non-detection in our observations means that the transit must have happened at least 3.9\,h too early or 4.5\,h too late. To give some constraints on the orbital period, we analysed the \textit{CoRoT} observations. To determine the range of periods that is excluded by our observations we carried out two individual weighted linear fits, one with the earliest possible transit mid-time after our observed window (transit 4.5\,h too late in respect to the original ephemeris) and the other with the latest possible transit mid-time before our observations (transit 3.9\,h too early). Equations \ref{Elemente_CoRoT27_long} and \ref{Elemente_CoRoT27_short} give a lower limit for a longer period and an upper limit for a shorter period, respectively.
\begin{equation}
\label{Elemente_CoRoT27_long}
\begin{array}{r@{.}lcr@{.}l}
T_{\mathrm{c[BJD_{TDB}]}}(E)=(2455748 & 6810 & + & E\cdot 3 & 575712)\,\mathrm{d}
\end{array}
\end{equation}
\begin{equation}
\label{Elemente_CoRoT27_short}
\begin{array}{r@{.}lcr@{.}l}
T_{\mathrm{c[BJD_{TDB}]}}(E)=(2455748 & 6905 & + & E\cdot 3 & 575004)\,\mathrm{d}
\end{array}
\end{equation}
Hence, we can exclude periods between 3.575004\,d and 3.575712\,d with our observations. However, the $\chi^{2}$ values of 37.8 and 5.3 for equation \ref{Elemente_CoRoT27_long} and equation \ref{Elemente_CoRoT27_short}, respectively, suggest, that a shorter period might be more likely. Therefore, photometric monitoring of CoRoT-27 a few hours before the predicted transit window is essential to recover the passage of this very interesting exoplanet in front of its host star. \\ As we could not add new transit events and, hence could not add new information, we did not re-determine the physical properties of CoRoT-27.

\section{Summary and Conclusions}

\begin{figure}
  \centering
  \includegraphics[width=0.33\textwidth,angle=270]{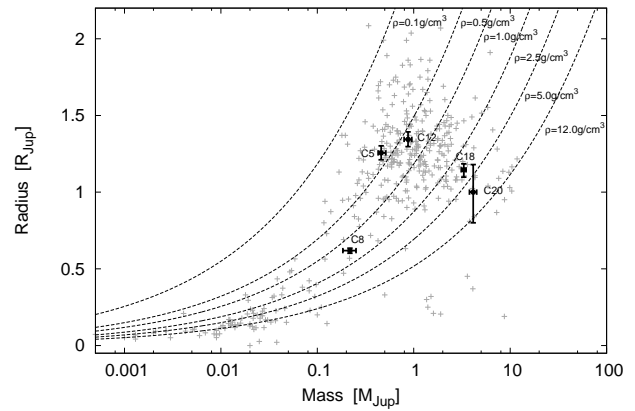}
  \caption{Mass-Radius diagram for transiting exoplanets, with our results for CoRoT-5\,b, CoRoT-8\,b, CoRoT-12\,b, CoRoT-18\,b, and CoRoT-20\,b. The lines of constant density (dashed lines) are also given.}
  \label{Mass_Radius}
\end{figure}

In our project to follow-up with ground-based photometry transiting planets discovered by the \textit{CoRoT} space telescope, we observed five systems between 2012 and 2016. The aim of our investigation has been to refine their orbital elements, constrain their physical parameters and search for additional bodies in the system. CoRoT-5, CoRoT-8, CoRoT-12, CoRoT-18, CoRoT-20, and CoRoT-27 were selected on the basis of their observability and expected photometric precision, of their at least slightly non-zero eccentricity (or little information to constrain the eccentricity) and /or of the uncertainties on their original published ephemeris.\\ Since \textit{CoRoT} could observe transiting planets continuously only for a maximum duration of 150\,d, the observations of our selected targets are well suited for our objectives because, on average, they took place 7\,yr after the exoplanet discovery. In total, we observed 14 transit events for five out of six targets. Despite the observation of two high precision LCs, we could not detect the expected transit of CoRoT-27\,b. \\ To conduct a homogeneous analysis of all available transit LCs, we re-analysed the observations of \textit{CoRoT}. We extracted all transit events, normalised, and cleaned (outlier removal) the LCs. With a total of 34, 25, 50, 17, and five transits for CoRoT-5\,b, CoRoT-8\,b, CoRoT-12\,b, CoRoT-18\,b, and CoRoT-20\,b, respectively, we performed simultaneous transit fitting in order to determine the system parameters. These were then used to calculate stellar, planetary and geometrical parameters of the systems. Our results for CoRoT-5\,b, CoRoT-8\,b, CoRoT-12\,b, CoRoT-18\,b, and CoRoT-20\,b plotted in a mass-radius diagram for transiting exoplanets are shown in Fig.~\ref{Mass_Radius}. CoRoT-5\,b is the planet with the lowest density, and CoRoT-8\,b the one with the lowest mass and radius in our sample. Approximately 70\% of the Jupiter-like transiting exoplanets (M$>$0.5M$_{\mathrm{Jup}}$) have a density between 0.2 and 1.2 $\mathrm{\rho}_{\mathrm{Jup}}$. Therefore, CoRoT-5\,b, CoRoT-8\,b, and CoRoT-12\,b have a comparable density to the majority of the transiting planets. CoRoT-18\,b and CoRoT-20\,b show a higher density. Only 8\% of the Jupiter-like transiting exoplanets have a higher density than CoRoT-20\,b. Hence, our measurements confirm that CoRoT-20\,b is one of the most compact planets known so far. \\ In most cases, our determined physical properties are in agreement with values reported in previous studies. For CoRoT-5, we found that the geometrical parameters are in excellent agreement, while the stellar and planetary values agree within the error bars on a 2$\sigma$ level. Also for CoRoT-12 and CoRoT-20, we found our derived physical properties in excellent (average deviation $\sim$0.5$\sigma$) and in good agreement (average deviation $\sim$1.0$\sigma$), respectively, with a largest deviation of 1.25$\sigma$. Only for CoRoT-8 and CoRoT-18, we found slight deviations from the literature values. In the case of CoRoT-18 this most likely arises from the different treatment of the stellar activity. For CoRoT-8 our LC derived stellar density is significant lower (on a 5-$\sigma$ level). The reason for these discrepancies were found to be strong parameter correlations in our LC modelling, which implies, e.g., that a smaller radius can be accounted for with a higher inclination $i$ without degrading the quality of the fit. By using a prior on the stellar density we derived physical properties that are in good agreement with the literature values. More high precision follow-up observations would be needed to break the degeneracies between the parameters.\\ In five out of six cases the observed mid-transit times deviate from the expected values more than estimated from the uncertainties on the original published ephemeris. One explanation is that the short observational baseline of \textit{CoRoT} does not allow for a precise determination of the orbital elements, and therefore the uncertainties on the original ephemeris were underestimated. For the CoRoT-27 system we could not even recover the transit event in the observing window predicted by the published ephemeris. The non-detection in our observations means that the transit must have happened at least 3.9\,h earlier or 4.5\,h later. Our analysis of the \textit{CoRoT}-transits suggests that the orbital period might be shorter than the literature one.  Hence, the confirmation of our finding would require to re-observe the system a few hours before the original transit time predictions. In the five remaining systems, CoRoT-5, CoRoT-8, CoRoT-12, CoRoT-18 and CoRoT-20, our re-determination of the orbital periods resulted in values that are between 0.9 and 29\,s longer and between 1.2 and 33 times more precise than the literature periods. Although some systems show a correlated structure of their transit times, we could not find significant periodicities in the timing residuals (FAP $\sim$99\% in all cases). A structured O--C diagram may also be caused by stellar activity. \\ Our ground-based photometric follow-up observations have allowed us to improve  the transit time predictions for six targets. In the era of space-based exoplanet characterization, accurate transit times are imperative for an efficient use of the observing time of future missions, like CHEOPS or JWST.  

\section*{Acknowledgements}

We would like to thank H. Gilbert for participating in some of the observations at the University Observatory Jena. \\ SR acknowledge support from the People Programme (Marie Curie Actions) of the European Union's Seventh Framework Programme (FP7/2007-2013) under REA grant agreement no. [609305]. MF acknowledges financial support from grants AYA2014-54348-C3-1-R, AYA2011-30147-C03-01 and AYA2016-79425-C3-3-P of the Spanish Ministry of Economy and Competivity (MINECO), co-funded with EU FEDER funds. CM acknowledges support from the DFG through grant SCHR665/7-1. The present study was made possible thanks to observations obtained with \textit{CoRoT}, a space project operated by the French Space Agency, CNES, with participation of the Science Program of ESA, ESTEC/RSSD, Austria, Belgium, Brazil, Germany, and Spain. \\ This research was (partly) based on data obtained at the 1.5m telescope of the Sierra Nevada Observatory (Spain), which is operated by the Consejo Superior de Investigaciones Cient\'{\i}ficas (CSIC) through the Instituto de Astrof\'{\i}sica de Andaluc\'{\i}a.

\bibliographystyle{mn2e}
\bibliography{literatur}


\label{lastpage}

\end{document}